\documentclass[12pt,preprint]{aastex}
\usepackage{lscape}

\newcommand{\msun}{{\rm \ M_\odot}}
\newcommand{\bdv}[1]{\mbox{\boldmath$#1$}}

\newcommand{\hei}{}
\def\rel{{\rm rel}}
\def\e{{\rm E}}

\def\kpc{{\rm kpc}}

\def\rel{{\rm rel}}
\def\max{{\rm max}}
\def\min{{\rm min}}
\def\e{{\rm E}}

\def\eff{{\rm eff}}

\begin{document}
\title{Planetary Detection Efficiency of the Magnification 3000 Microlensing
Event OGLE-2004-BLG-343}

\author{Subo Dong\altaffilmark{1},
        D.L. DePoy\altaffilmark{1},
        B.S. Gaudi\altaffilmark{2},
        A. Gould\altaffilmark{1},
        C. Han\altaffilmark{3},
        B.-G. Park\altaffilmark{4},
        R.W. Pogge\altaffilmark{1} \\
        (The $\mu$FUN Collaboration), \\
        A. Udalski\altaffilmark{5},
        O. Szewczyk\altaffilmark{5},
        M. Kubiak\altaffilmark{5},
        M.K. Szyma{\'n}ski\altaffilmark{5},
        G. Pietrzy{\'n}ski\altaffilmark{5,6},
        I. Soszy{\'n}ski\altaffilmark{5,6},
        {\L}. Wyrzykowski\altaffilmark{5},
        K. {\.Z}ebru{\'n}\altaffilmark{5} \\
        (The OGLE Collaboration) \\
}
\affil{
\altaffiltext{1}
{Department of Astronomy, Ohio State University,
140 W.\ 18th Ave., Columbus, OH 43210, USA; dong, depoy, gould, pogge@astronomy.ohio-state.edu}
\altaffiltext{2}
{Harvard-Smithsonian Center for Astrophysics, 
Cambridge, MA 02138, USA; sgaudi@cfa.harvard.edu}
\altaffiltext{3}
{Department of Physics, Institute for Basic Science Research,
Chungbuk National University, Chongju 361-763, Korea;
cheongho@astroph.chungbuk.ac.kr}
\altaffiltext{4}
{Korea Astronomy and Space Science Institute, Daejeon 305-348, Korea; 
bgpark@kasi.re.kr}
\altaffiltext{5}
{Warsaw University Observatory, Al.~Ujazdowskie~4, 00-478~Warszawa, Poland;
udalski, szewczyk, mk, msz, pietrzyn, soszynsk, wyrzykow, zebrun@astrouw.edu.pl
}
\altaffiltext{6}
{Universidad de Concepci{\'o}n, Departamento de Fisica,
Casilla
160--C, Concepci{\'o}n, Chile}
}
\begin{abstract}
OGLE-2004-BLG-343 was a microlensing event with peak
magnification $A_{\rm max}=3000\pm 1100$, 
by far the highest-magnification event ever
analyzed and hence potentially extremely sensitive to planets
orbiting the lens star.  Due to human error, intensive
monitoring did not begin until 43 minutes after peak, at which point
the magnification had fallen to $A\sim 1200$, still by far the
highest ever observed.  As the light curve does not show significant
deviations due to a planet, we place upper limits on the presence
of such planets by extending the method of \citet{yoo423}, which combines
light-curve analysis with priors from a Galactic model of the source
and lens populations, to take account
of finite-source effects.  This is the first event so analyzed for
which finite-source effects are important, and hence we develop two new
techniques for evaluating these effects.  Somewhat surprisingly,
we find that OGLE-2004-BLG-343 is no more sensitive to planets than
two previously analyzed events with $A_{\rm max}\sim 100$, despite
the fact that it was observed at $\sim 12$ times higher magnification.
However, we show that had the event been observed over its peak, it
would have been sensitive to almost all Neptune-mass planets 
over a factor of 5 of 
projected separation and even would have had some sensitivity to Earth-mass
 planets. This shows that 
some microlensing events being detected 
in current experiments are sensitive to very low-mass planets. 
We also give suggestions on how extremely high-magnification
events can be more promptly monitored in the future.
\end{abstract}


\section{Introduction
\label{intro}}

{\hei Current microlensing planet searches focus significant effort on 
high-magnification events, which have great promise for
detecting low-mass extrasolar planets. It is therefore crucial 
to understand the potential for discovering planets  
and to optimize the early identifications and observational strategy
of such events. 
In previous planetary detection efficiency analyses of 
high-magnification events, finite-source effects have 
often been ignored mainly due to computational 
limitations. However, such effects are intrinsically 
important in these events because the sources are more likely to be 
resolved at very low impact parameters. In this study, we improve the 
method of \citet{yoo423} by incorporating finite-source effects
to characterize the planetary detection efficiency of the 
extremely high-magnification event OGLE-2004-BLG-343, and
we develop new efficient algorithms to make the calculations
possible. Moreover, we attempt to find useful observational signatures of 
high-magnification events so as to help alleviate the difficulties in 
their early recognition.

\subsection{High-Magnification Events \& Earth-Mass Planets}
\label{sec:intro1}
}

Apart from pulsar timing {\hei \citep{pulsar}}, microlensing is at present 
one of the few planet-finding techniques that is sensitive to 
Earth-mass planets.
A planetary companion of an otherwise isolated lens star introduces
two kinds of caustics into the magnification pattern: ``planetary caustics''
associated with the planet itself and a ``central caustic''
associated with the primary lens. \footnote{When the 
planetary companion is close to the Einstein ring, 
the planetary and central caustics merge into 
a single ``resonant caustic''.}
When the source passes over or
close to one of these caustics, the light curve deviates from
its standard \citet{pac86} form, thus revealing the presence
of the planet \citep{mao91,liebes}.

Since planetary caustics are generally far larger than central
caustics, a ``fair sample'' of planetary microlensing events would
be completely dominated by planetary-caustic events.  Nevertheless,
central caustics play a crucial role in current microlensing planet
searches, particularly for Earth-mass planets \citep{griest98}, 
for the simple reason that it is possible to predict in advance 
that the source of a given
event will arrive close to the center of the magnification pattern
where it will probe for the presence of these caustics.  Hence,
one can organize the
intensive observations required to characterize the resulting
anomalies.  By contrast, the perturbations due to
planetary caustics occur without any warning.  The lower the
mass of the planet, the shorter the duration of the anomaly,
and so the more crucial is the warning to intensify the observations.
This is the primary reason that planet-searching groups give
high priority to high-magnification events, i.e., those that probe
the central caustics \footnote{\hei Note that although
high-magnification events are guaranteed to have low impact parameters,
peak magnification for events with low impact parameters are not necessarily
high if they have relatively large source sizes. And large sources will tend to
smear out the perturbations induced by the central caustics, thereby
decreasing the planetary sensitivity \citep{griest98, han05}.
So when central caustics are important in producing planetary signals,
the maximum magnification of microlensing events serves as a better
indicator of planetary detection efficiency than the impact parameter.
}. As a bonus, high-magnification events
are also more sensitive to planetary-caustic perturbations than are
typical events \citep{gouldloeb92} {\hei since their larger images  
increase the chances that they will be perturbed by planets}.  
However, this enhancement is
relatively modest compared to the rich potential of central-caustic
crossings.

In principle, it is also possible to search for Earth-mass planets from
perturbations due to their larger (and so more common) planetary and resonant 
caustics, but this would require a very different strategy from those 
currently being carried out. The problem is that these perturbations occur 
{\it without warning} during otherwise
normal microlensing events, and typically last only 1 or 2 hr.
Hence, one would have to intensively monitor the entire duration of many
events. The only way to do this practically is to intensively monitor
an entire field containing
many ongoing microlensing events roughly once every 10 minutes in order
to detect and properly characterize the planetary deviations.  Proposals
to make such searches have been advanced for both space-based
\citep{bennettrhie02} and ground-based \citep{sackett97}
platforms.

At present, two other microlensing planet-search strategies are being
pursued.  Both strategies make use of wide-area $(\ga 10\,\rm deg^2)$
searches for microlensing events toward the Galactic bulge.  Observations
are made once or a few times per night by the OGLE-III\footnote{
OGLE Early Warning System: http://www.astrouw.edu.pl/\~{}ogle/ogle3/ews/ews.html}
\citep{udal03} and MOA\footnote{MOA Transient Alert Page: 
http://www.massey.ac.nz/\~{}iabond/alert/alert.html}
\citep{bond01} surveys.  When events are identified, they are posted
as ``alerts'' on their respective Web sites.
In the first approach, these groups check each ongoing event after
each observation for signs of anomalous behavior, and if their
instantaneous analysis indicates that it is worth doing so, they switch
from survey mode to follow-up mode. This approach led to 
the first reliable detection of a planetary
microlensing event, OGLE 2003-BLG-235/MOA-2003-BLG-53 \citep{bond04}.

In the second approach, follow-up groups such as the Probing Lensing Anomalies
NETwork (PLANET; \citealt{albrow98})
and the Microlensing Follow-Up Network ($\mu$FUN; \citealt{yoo423})
monitor a subset of alerted events many times per day and from locations
around the globe.  Generally these groups focus to the extent possible
on high-magnification events for the reasons stated above. The survey groups
can also switch from ``survey mode'' to ``follow-up'' mode to probe newly
emerging high-magnification events.

Over the past decade several high-magnification events have been analyzed
for planets.  \citet{gaudi02} and \citet{albrow01} placed upper limits on 
the frequency of
planets from the analysis of 43 microlensing events, three of which
reached magnifications $A_{\rm max}\geq 100$, including
OGLE-1998-BUL-15 ($A_{\rm max}=170\pm 30$),
MACHO-1998-BLG-35 ($A_{\rm max}=100\pm 5$), and
OGLE-1999-BUL-35 ($A_{\rm max}=125\pm 15$).  
However, the first of these was not monitored over its peak.
MACHO-1998-BLG-35 was also analyzed by \citet{rhie00} and \citet{bond02},
who incorporated all available data and found modest $(\Delta\chi^2=63)$ 
evidence for one, or perhaps two, Earth-mass planets.

\citet{yoo423} analyzed OGLE-2003-BLG-423 ($A_{\rm max}=256\pm 43$), which
at the time was the highest magnification event yet recorded.  However,
because this event was covered only intermittently over the peak,
it proved less sensitive to planets than either MACHO-1998-BLG-35 or
OGLE-1999-BUL-35.

\citet{abe04} analyzed MOA-2003-BLG-32/OGLE-2003-BLG-219, which at
$A_{\rm max}=525\pm 75$, is the current record-holder for maximum
magnification.  Unlike  OGLE-2003-BLG-423, this event was monitored
intensively over the peak: the Wise Observatory in Israel was able
to cover the entire 2.5 hr FWHM during
the very brief interval that the bulge is visible from this northern
site.  The result is that this event has the best sensitivity
to low-mass planets to date.

Recently, \citet{ob05071} detected a $\sim 3$-Jupiter mass
planet by intensively monitoring the peak of the high-magnification
event OGLE-2005-BLG-071.  This was the second robust detection
of a planet by microlensing and the first from perturbations
due to a central caustic.

\subsection{Planet Detection Efficiencies: Philosophy and Methods}
\label{sec:subintro}

{\hei The fundamental aim of microlensing planet searches is to 
derive meaningful conclusions about the presence of
planets (or lack thereof) from these searches. Therefore, it is essential
to quantitatively assess what planets could have been detected
from the observations of individual non-planetary
events if such planets had
been present.}  Actually, this problem is not as easy to properly
formulate as it might first appear.  For example, 
the event parameters are measured with only finite precision.
Among these, the impact parameter $u_0$ (in units of the angular Einstein
radius $\theta_\e$) is {\hei particularly important: if the event really}
did have a $u_0$ equal to its best-fit value, then one could 
calculate whether
a planet at a certain separation and position angle would have
given rise to a detectable signal in the observed light curve.
But the true value of $u_0$ may differ from the best-fit value
by, say, $1\,\sigma$, and the same planet may not give rise to
a detectable signal for this other, quite plausible geometry.
{\hei (In principle, an error in the time of maximum, $t_0$,
would cause a similar problem, displacing the assumed path through
the Einstein ring by $\delta t_0/t_{\rm E}$, where $t_{\rm E}$
is the Einstein crossing time.  However, because
$u_0$ is strongly correlated with several other parameters
while $t_0$ is not, the error in $u_0$ is substantially
larger than the error in $t_0$ divided by $t_{\rm E}$.)}
Or, as another example, consider finite-source effects.  
Planetary perturbations have a fairly high probability of exhibiting
finite-source effects, which then have a substantial impact
on whether the deviation can be detected in a given data stream.
If there is such a planetary perturbation, one can measure 
$\rho_*=\theta_*/\theta_\e$, the size of the source relative to the
Einstein radius.  But if there is no planet detected, no finite-source
effects are typically detected, and hence there is no direct information
on $\rho_*$.  Therefore, one cannot reliably determine whether a given
planetary perturbation would have been affected by finite-source
effects and so whether it would have been detected.  Finally, there
are technical questions as to what exactly it means that a planet
``would have been'' detected.

The past decade of microlensing searches has been accompanied by a 
steady improvement in our understanding of these questions.
\citet{gaudi00} developed the first method to evaluate
detection efficiencies, which was later implemented by \citet{albrow00}
and \citet{gaudi02}.  In this approach, binary models are fitted to the
observed data with the three ``binary parameters'' $(b,q,\alpha)$
held fixed and the three ``point-lens parameters'' $(t_0,u_0,t_\e)$
allowed to vary.  Here $b$ is the planet-lens separation in units
of $\theta_\e$, $q$ is the planet-star mass ratio, $\alpha$ is the angle of 
the source trajectory relative to the binary axis, $t_0$
is the time of the source's closest approach to the center of the
binary system, $u_0=u(t_0)$ is the impact parameter, $t_E=\theta_\e/\mu$
is the Einstein timescale, and $\mu$ is the source-lens relative proper motion.
If a particular $(b,q,\alpha)$ yielded a $\chi^2$ improvement 
$\Delta\chi^2<\chi^2_{\rm min}=-60$, a planet could be said to be detected.
If not, then the ensemble of $(b,q,\alpha)$ for which 
$\Delta\chi^2>\chi^2_{\rm min}=60$ was said to be excluded for that
event.  For each $(b,q)$, the fraction of angles $0\leq\alpha\leq 2\pi$
that was excluded was designated the ``sensitivity'' for that geometry.

\citet{gaudi02} argued that this method underestimated the sensitivity
because it allowed the fit to move $u_0$ to values for which the
source trajectory would ``avoid'' the planetary perturbation
but still be consistent with the light curve.  That is, $u_0$ 
has some definite value, even if it were not known to
the modelers exactly what that value should be. \citet{yoo423} followed up
on this by holding $u_0$ fixed at a series of values and estimated
planetary detection efficiency at each. The total efficiency would then be
the average of these weighted by the probability of each value of
$u_0$.  In principle, one should also integrate over $t_0$ and
$t_\e$.  In practice, \citet{yoo423} found that, at least for the
event they analyzed, $t_0$ and $t_{\rm eff}\equiv u_0 t_\e$ were
determined very well by the data, so that once $u_0$ was fixed,
so were $t_0$ and $t_\e$.

\citet{yoo423} departed from all previous planet-sensitivity estimates
by incorporating a Bayesian analysis that accounts for priors 
derived from a Galactic model of the mass, distance, and velocity 
properties of source and lens population into the analysis. They 
simulated an ensemble of events and weighted each by {\it both} the
prior probability of the various Galactic model parameters 
(lens mass, lens and source distances, lens and source velocities)
and the goodness of fit of the resulting magnification profile to the
observed light curve.  This approach was essential to enable a
proper weighting of different permitted values of $u_0$.  As a bonus,
it allowed one, for the first time, to determine the sensitivities
as a function of the physical planetary parameters (such as planet mass
$m_p$ and planet-star separation $r_\perp$) as opposed to the
microlensing parameters, the planet-star mass ratio $q$ and the planet-star
projected separation (in the units of $\theta_\e$) $b$.

\citet{rhie00} introduced a procedure for evaluating planet 
sensitivities that differs qualitatively from that of 
\citet{gaudi00}.  For each trial $(b,q,\alpha)$ and observed
point-lens parameters $(t_0,u_0,t_\e)$, they created a simulated
light curve with epochs and errors similar to those of the real 
light curve.  They then fitted this light curve to a point-lens model
with $(t_0,u_0,t_\e)$ left as free parameters.  If the point-lens
model had $\Delta\chi^2>\chi^2_{\rm min}$, then this $(b,q,\alpha)$
combination was regarded as excluded.  That is, they 
mimicked their planet-detection procedure on simulated planetary events.

\citet{abe04} carried out a similar procedure except that they
did not fit for $(t_0,u_0,t_\e)$, but rather just held these
three parameters fixed at their point-lens-fit values.  Of course,
this procedure necessarily yields a higher $\Delta\chi^2$ than
that of \citet{rhie00}, but \citet{abe04} expected that the
difference would be small.

While all workers in this field have recognized that finite-source
effects are important in principle, they have generally concluded
that these did not play a major role in the particular events that
they analyzed.  This has proved fortunate because the source size
is generally unknown, and even a single trial value for the source
size typically requires several orders of magnitude more computing
time than does a point-source model.  \citet{gaudi02} estimated
angular sizes $\theta_*$ of each of their 43 source stars from their 
positions on an instrumental color-magnitude diagram (CMD) by adopting
$\mu=12.5\,\rm km\,s^{-1}\,kpc^{-1}$ for all events and evaluating
$\rho_* = \theta_*/(\mu t_\e)$.  They made their sensitivity estimates
for both this value of $\rho_*$ and for a point source ($\rho_*=0$)
and found that generally the differences were small.  They concluded that
a more detailed finite-source evaluation was unwarranted (and also
computationally prohibitive).  Using their Monte Carlo technique,
\citet{yoo423} were able to evaluate the probability distribution of
the parameter combination {\hei $z_0$, which is equal to impact 
parameter over source size}.  This analysis
showed that $z_0\gg 1$ {\hei(no finite-source effects)} 
with high confidence for their event.  This implied that
the source did not pass close to the central caustic and hence
that finite-source effects were not important.  
Again, computation
for additional values of $\rho_*$ would have been computationally
prohibitive.

{\hei
\subsection{"Seeing" the Lens in High-Magnification Events}
\label{sec:lense}

In the very first paper on microlensing, \citet{einstein36} already
realized that it might be difficult to observe the magnified source due to
``dazzling by the light of the much nearer [lens] star.''

Seventy years later, more than 2000 microlensing events have been
discovered, but only for two of these has the ``dazzling'' light of the
lens star been definitively observed. The best case is MACHO-LMC-5,
for which the lens was directly imaged by the
{\it Hubble Space Telescope (HST)} \citep{alcock01,drake04}, which
yielded mass and distance estimates of the lens that agreed to good
precision with those derived from the microlensing event itself
\citep{gba04}.

The next best case is OGLE-2003-BLG-175/MOA-2003-BLG-45, for which
\citet{ghosh04} showed that the blended light was essentially
perfectly aligned with the source. This would be expected if the blend
actually was the lens, but it would be most improbable if it were just an
ambient field star.  In this case, the blend was about 1 mag brighter than
the baseline source in $I$ and 2 mag brighter in $V$, perhaps
fitting Einstein's criterion of a ``dazzling'' presence.

Intriguingly, the above two events positively identified to harbor
luminous lenses are both high-magnification. It is quite plausible that
events with luminous lenses are biased towards high magnification since
they will more likely be missed if the magnifications are too low. This raises
the question of whether OGLE-2003-BLG-423 has a luminous lens. In addition,
identifying the lens star would allow us to directly determine the
 physical properties of the lens, which in turn would help better 
constrain parameters in the planetary detection analysis.
}

Here we analyze OGLE-2004-BLG-343, whose maximum magnification 
$A_{\rm max}\sim 3000$ is by far the highest of any observed event and
the first to exceed the $A = 1000$ benchmark initially discussed by 
\citet{liebes} as roughly the maximum possible magnification for typical
Galactic sources and lenses.\footnote{\citet{liebes} derived
that for perfect lens-source alignment, $A_{u=0}=2{\theta_\e/\theta_*}$ by
approximating the source star as having uniform surface brightness, and
he evaluated this expression for several typical examples.}

As we describe below, the event was alerted as a possibly anomalous,
very-high-magnification event in time to trigger
intensive observations over the peak, but due to human error,
the actual observations caught only the falling side of the peak.
We analyze both the actual observations made of this event (in
order to evaluate its actual sensitivity to planets) and the
sequence of observations that should have been initiated by
the trigger.  The latter calculation illustrates the
potential of state-of-the-art microlensing observations, although
unfortunately this potential was not realized in this case.

We analyze the event within the context of the \citet{yoo423} formalism
with several major modifications.  First, we adopt the \citet{rhie00}
criterion of planet-sensitivity in place of that of \citet{gaudi00}.
That is, we say a planet configuration is ruled out if simulated data
generated by this configuration are inconsistent with
a point-lens light curve at $\Delta\chi^2>\Delta\chi^2_{\rm min}$
.  Second, using a Monte Carlo
simulation, we show that for this event, $z_0\sim 1$, and hence 
finite-source effects are quite important.  This requires us to
generalize the \citet{yoo423} procedure to include a two-dimensional
grid of trial parameters $(u_0,\rho_*)$ in place of the one-dimensional $u_0$
sequence used by \citet{yoo423}.  
From what was said above, it
should be clear that the required computations would be completely
prohibitive if they were carried out using previous numerical 
techniques.  Therefore, third, we develop new techniques for finite-source
calculations that are substantially more efficient than those used previously.

In \S~\ref{sec:data}, we describe the data. Next, we discuss
modeling of the event in \S~\ref{sec:model}. Then in \S~\ref{sec:dp},
we present our procedures and results related to planet detection. 
We explore the possibility that the blended light is due to the 
lens in \S~\ref{sec:luminouslens}. In \S~\ref{sec:summary} we summarize 
our results and make suggestions on monitoring extremely high-magnification
events in the future. Finally, 
the two new binary-lens finite-source algorithms that we have developed are 
described in Appendix~\ref{sec:code}.

\section{Observational Data
\label{sec:data}}
The first alert on OGLE-2004-BLG-343 was triggered by the OGLE-III Early Warning 
System (EWS; \citealt{udal03}) on 2004 June 16, about 3 days before its peak 
on HJD$'\equiv$ HJD$-2450000 = 3175.7467$. 
On June 18, after the first observation of the lens was taken, 
the OGLE real time lens monitoring system 
(Early Early Warning System [EEWS]; \citealt{udal03}) triggered an internal
alert, indicating a deviation from the single lens light curve based on
previous data. Two additional observations were made after that, but the new fits to all of the data  were still fully compatible with single-mass lens albeit suggesting high magnification at maximum. Therefore, an alert
to the microlensing community was distributed by OGLE on HJD$'$ = 3175.1 suggesting OGLE-2004-BLG-343 as a possible high-magnification event.
The observation at UT 0:57 (HJD$'$ 3175.54508) the next night showed 
a large deviation from the light-curve prediction based on previous observations, and 
an internal EEWS alert was triggered again. Usually further observations would have been 
made soon after such an alert, but unfortunately no observations were performed 
until UT 6:29 (HJD$'$ 3175.77626), about 0.71 hr after the peak. At that 
time, the event had brightened by almost 3 mag in $I$ relative to the previous night's 
observation, and therefore it was regarded very likely to be the first caustic 
crossing of a binary-lens event. As a consequence, no $V$-band photometry was
undertaken to save observation time and in the hope that observations in $V$
could be done when it brightened again. {\hei  
After two post-peak observations confirmed the event's
extremely high magnification, OGLE began maximally
intensive observations with a cadence of 4.3 minutes.
However,
after it was clear that the event was falling in a regular fashion, it
 was then observed less intensively. 
A total of 14 observations were performed during 3.39 hr, and a new alert to 
the microlensing community was immediately released by OGLE as well.}
However, the next day the event faded drastically, by 
about 3 mag from the maximum point of the previous night, implying that if the
event were a binary, the peak had probably been a cusp crossing rather than a 
caustic crossing. After being monitored for a few more days, it became clearer 
that OGLE-2004-BLG-343 was most probably a point-lens event of very high magnification
and therefore very sensitive to planets. This recognition prompted OGLE to
obtain a $V$-band point, but by this time (HJD$'$ 3179.51) the source had fallen
6 mag from its peak, so that only a weak detection of the $V$ flux was possible. Hence,
this yielded only a lower limit on the $V-I$ color. 

By chance, $\mu$FUN made one dual-band observation in $I$
and $H$ 1 day before peak (HJD$'$ 3174.74256) solely as a reference point to
check on the future progress of the event. After the event peaked, $\mu$FUN also concluded
that it was uninteresting until OGLE/$\mu$FUN email exchanges led to the conclusion
that the event was important. Since the source was magnified by A$\sim$40 at this pre-peak
 $\mu$FUN 
observation, it enabled a clear $H$-band detection and so yielded an $(I-H)$ color measurement, 
which can be translated to $(V-I)$.

The OGLE data are available at the OGLE EWS Web site mentioned above
and the $\mu$FUN data are available at the $\mu$FUN Web site
\footnote{http://www.astronomy.ohio-state.edu/\~{}microfun}.

There were 195 images in $I$ and eight images in $V$, both from OGLE, as well
as three images in $H$ from $\mu$FUN.
Since only OGLE $I$-band observational data are available near the
peak, the following analysis is entirely based on the OGLE $I$-band data except that the 
OGLE $V$-band data and $\mu$FUN $H$-band data are used to constrain the color of 
the source star. The OGLE errors
are renormalized by a factor of 1.42 so that the $\chi^2$ per degree of freedom for 
the best-fit point-source/point-lens (PSPL) model is close to unity. We also eliminate 
the two OGLE points that are $3\,\sigma$ outliers. These are both well away from the peak
{\hei and therefore their elimination has no practical impact on our analysis}. 

\section{Event Modeling
\label{sec:model}}
\citet{yoo423} introduced a new approach to model microlensing
events for which $u_0$ is not perfectly measured. As distinguished from previous 
analyses, this method establishes the prior probability of the event parameters by
performing a Monte-Carlo simulation of the event using a Galactic model
 rather than simply assuming uniform distributions. This approach is not only
more realistic but also makes possible the estimation of physical parameters, which are
otherwise completely degenerate. 
Following the procedures of \citet{yoo423}, we begin our modeling by fitting the event 
to a PSPL model, evaluating the 
finite-source effects and performing a Monte-Carlo simulation. We then improve that 
method by considering finite-source effects when combining the simulation with the light-curve fits. 

\subsection{Point-Source Point-Lens Model}
\label{sec:pspl}

The PSPL magnification is given by \citep{pac86}
\begin{equation}
A(u) = {u^2+2\over u\sqrt{u^2+4}},\qquad{u(t)} = \sqrt{u_0^2 + {{(t-t_0)^2}\over{t_\e^2}}},
\label{eqn:mag}
\end{equation}
where $u$ is the projected lens-source separation in units of the angular Einstein 
radius $\theta_\e$, $t_0$ is the time of maximum magnification, $u_0=u(t_0)$ is
the impact parameter, and $t_\e$ is the Einstein timescale.

The predicted flux is then
\begin{equation}
F(t) = F_{s}A[u(t)] + F_{b},
\label{eqn:flux}
\end{equation}
where $F_{\rm s}$ is the source flux and $F_{\rm b}$ is the blended-light flux.

\begin{figure}
\plotone{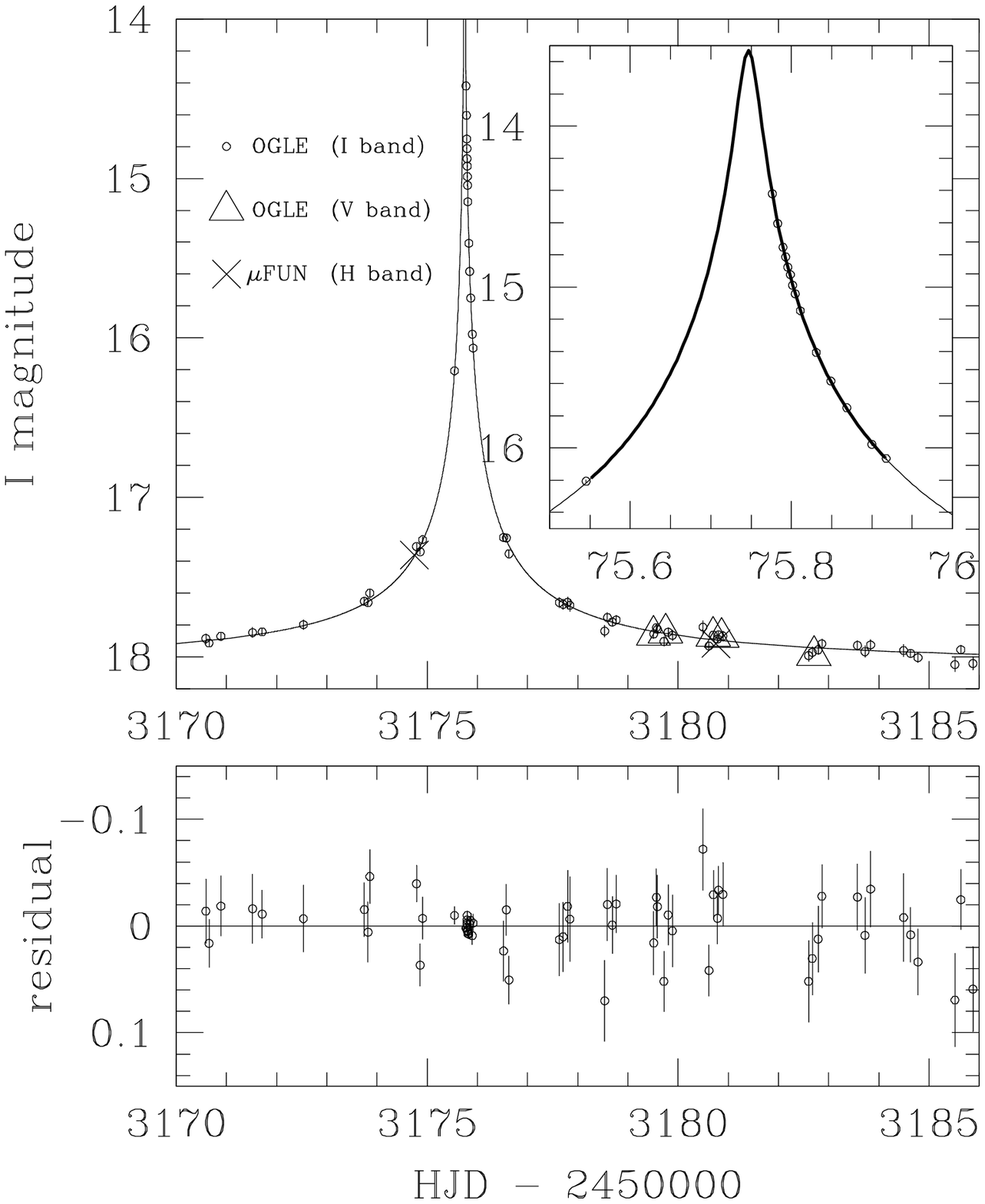}
\caption{\label{fig:lc}
Light curve of OGLE-2004-BLG-343 near its peak on 2004 June 19
(HJD 2,453,175.7467). Only OGLE $I$-band data ({\it open circles}) are used in
most of the analysis, except OGLE $V$-band data ({\it open triangles}) 
and $\mu$FUN $H$-band data ({\it crosses}) are used to constrain the color of the 
source star. All bands are linearly rescaled so that $F_{\rm s}$ and $F_{\rm b}$ are
the same as the OGLE $I$-band observations. The solid line shows the best-fit
PSPL model. The upper right inset shows the peak of the light curve, with the 
range of the simulated data points plotted by the thick line.}
\end{figure}

The observational data are fitted in the above model with five free parameters 
($t_0$, $u_0$, $t_\e$, $F_{\rm s}$, and $F_{\rm b}$). The results of the fit are shown
in Table~\ref{table1} (also see Fig.~\ref{fig:lc}). The best-fit $u_0$ is remarkably small, 
$u_0 = 0.000333 \pm 0.000121$, which indicates that the maximum magnification 
is $A_{\max} = 3000 \pm 1100$. As discussed below in \S~\ref{sec:fs}, 
the $3\sigma$ lower limit is $A_{\rm max} \ga 1450$.
This is the first microlensing event 
ever analyzed in the literature with peak magnification higher than 1000. 
The uncertainties in $u_0$, $t_E$ and $F_{\rm s}$ are fairly large, roughly 35\%. As pointed out
in \citet{yoo423}, these errors are correlated, while combinations of these
parameters, $t_{\eff} \equiv u_0 t_\e$ and $F_{\max}\equiv F_{\rm s}/u_0$, have much
smaller errors:
\begin{equation} 
t_\eff = 0.0141 \pm 0.0008  \,\rm days,
\qquad I_\min = 13.805 \pm 0.065.
\label{eqn:teffimin}
\end{equation}
Here $I_\min$ is the calibrated $I$-band magnitude corresponding to $F_\max$.

\subsection{Source Properties from Color-Magnitude Diagram}
\label{sec:cmd}

It is by now standard practice to determine the dereddened 
color and magnitude of a microlensed source by putting 
the best-fit instrumental color and magnitude of the source
on an instrumental $(I,V-I)$ CMD.  The
dereddened color and magnitude can then be determined from
the offset of the source position from the center of the
red clump, which is locally measured to be $\left[M_I,(V-I)_0\right] = (-0.20,1.00)$. We adopt a 
Galactocentric distance $R_0=8\,$kpc. However, at Galactic
longitude $l=+4.21$, the red clump stars in the OGLE-2004-BLG-343 field are closer to us than the
Galactic center by $0.15$ mag \citep{stanek97}. 
We derive $(I,V-I)_{0,\rm clump} = (14.17,1.00)$. Although the source instrumental 
color and magnitude are both fit parameters,
only the magnitude is generally strongly correlated with other
fit parameters.  By contrast, the source instrumental color
can usually be determined directly by a regression of $V$ on $I$ flux
as the magnification changes.  No model of the event is actually
required to make this color determination. In the present case, we exploit
both $(V-I)$ and $(I-H)$ data. Hence, in order to make use of this technique,
we must convert the $(I-H)$ to $(V-I)$. This will engender some difficulties.

\begin{figure}
\plotone{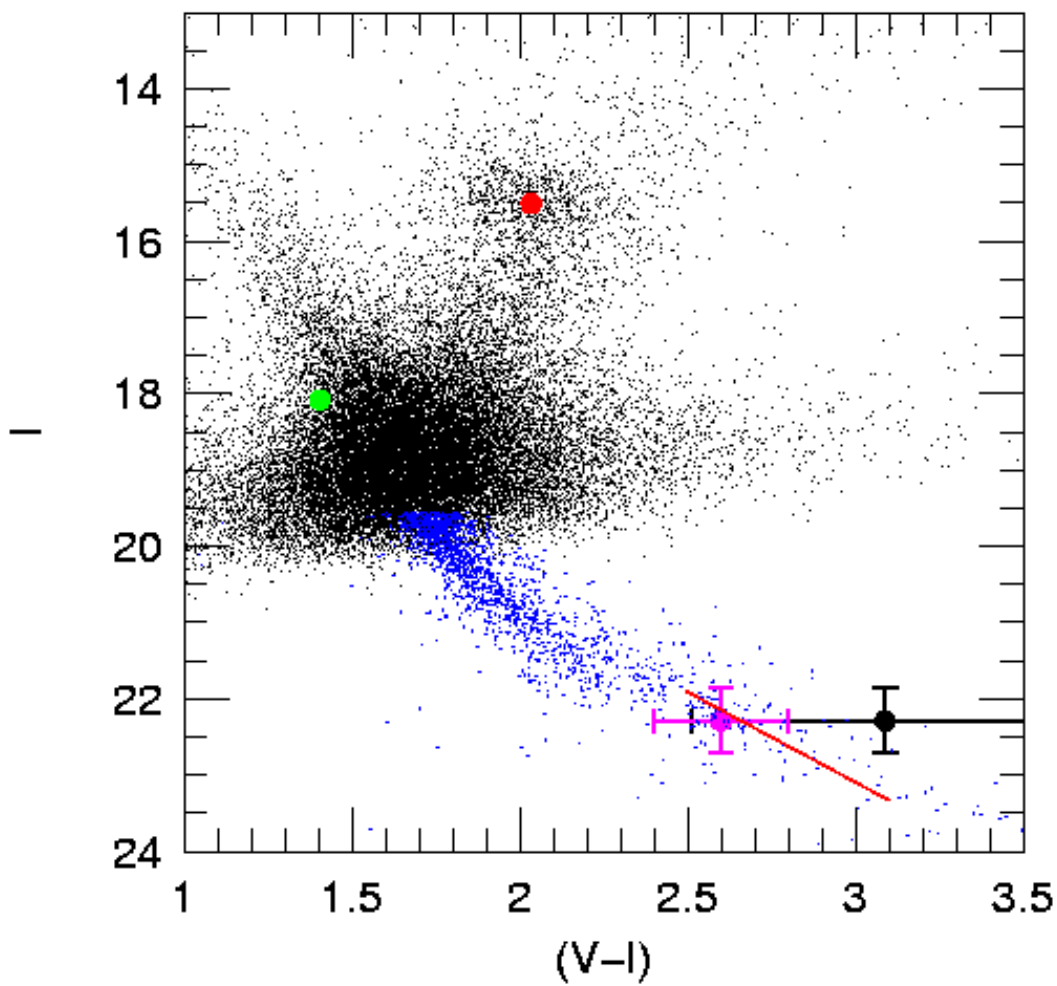}
\caption{
\label{fig:fcmd}
CMD of the OGLE-2004-BLG-343 field.
{\it Hipparcos} main-sequence stars ({\it blue dots}), placed at $10^{-0.15/5}R_0=7.5\,$kpc 
and reddened by the reddening vector derived from the clump, are displayed with
the OGLE-II stars ({\it black dots}). The \citet{reid91} relation is plotted 
by the red solid line over the {\it Hipparcos} stars. On the CMD, the magenta filled circle 
is the red clump and the green filled circle is the blended star. The large black filled circle
is the OGLE $V$  measurement of the source with of $1\sigma$ error bars, 
which sets a lower limit for the source $V-I$ color. The magenta filled circle with 
error bars is the result of combining the $(I-H)/(V-I)$ information 
(see Fig.~\ref{fig:vih}) with the OGLE measurement.
}
\end{figure}

As discussed in \S~\ref{sec:data}, however, $V$-band measurements were
begun only when the source had fallen nearly to baseline.  Hence,
the measurement of the $(V-I)$ color obtained by this standard
procedure has very large errors and indeed is consistent with
infinitely red ($F_{{\rm s},V}=0$) at the $2\,\sigma$ level
(see Fig.~\ref{fig:fcmd}).  The CMD itself is based on OGLE-II
photometry, and we have therefore shifted the OGLE-III-derived
fluxes by $\Delta I = I_{\rm OGLE-II} - I_{\rm OGLE-III} = 0.26$ mag.
On this now {\it calibrated} CMD, the clump is at
$(I,V-I)_{\rm clump} = (15.51,2.04)$.  
Hence, the dereddened source color and magnitude are given by
$(I,V-I)_0 = (I,V-I) + (I,V-I)_{0,\rm clump} - (I,V-I)_{\rm clump}
= (I,V-I) - (1.34,1.04)$, the final offset being the reddening vector.
This vector corresponds to $R_{VI} = 1 + 1.34/1.04 = 2.29$, which
is somewhat high compared to values obtained by \citet{sumi} for typical
bulge fields. However, {\hei we will present below independent evidence
 for this or a slightly higher value of $R_{VI}$}.  Figure~\ref{fig:fcmd} also
shows the position of the blended light, {\hei which lies in the so-called
reddening sequence of foreground disk main sequence stars}.
This raises the question as to whether this blended star is actually the lens. 
We return to this question in \S~\ref{sec:luminouslens}.

The source star is substantially fainter than any of the other stars
in the OGLE-II CMD.  In order to give a sense of the relation between
this source CMD position and those of main-sequence bulge stars, we also
display the {\it Hipparcos} main sequence \citep{hip}, placed at 
$10^{-0.15/5}R_0=7.5\,$kpc and
reddened by the reddening vector derived from the clump.  At the
best-fit value, $V-I=3.09$, the source lies well in front of (or to
the red of) the bulge main-sequence.  However, given the large color
error, it is consistent with lying on the bulge main sequence at the
$1\,\sigma$ level.

To obtain additional constraints on the color, we consider the $\mu$FUN
instrumental $H$-band data.
The single highly-magnified $(A\sim 40)$ $H$-band point (together
with a few baseline points) yields $I_{\rm OGLE-II}-H_{\rm {\mu}FUN} = 0.59 \pm 0.11$ 
source color. To be of use, this must be translated to a $(V-I)_{\rm OGLE-II}$
color using a $(V-I)/(I-H)$ color-color diagram of the stars in the field.

\begin{figure}
\plotone{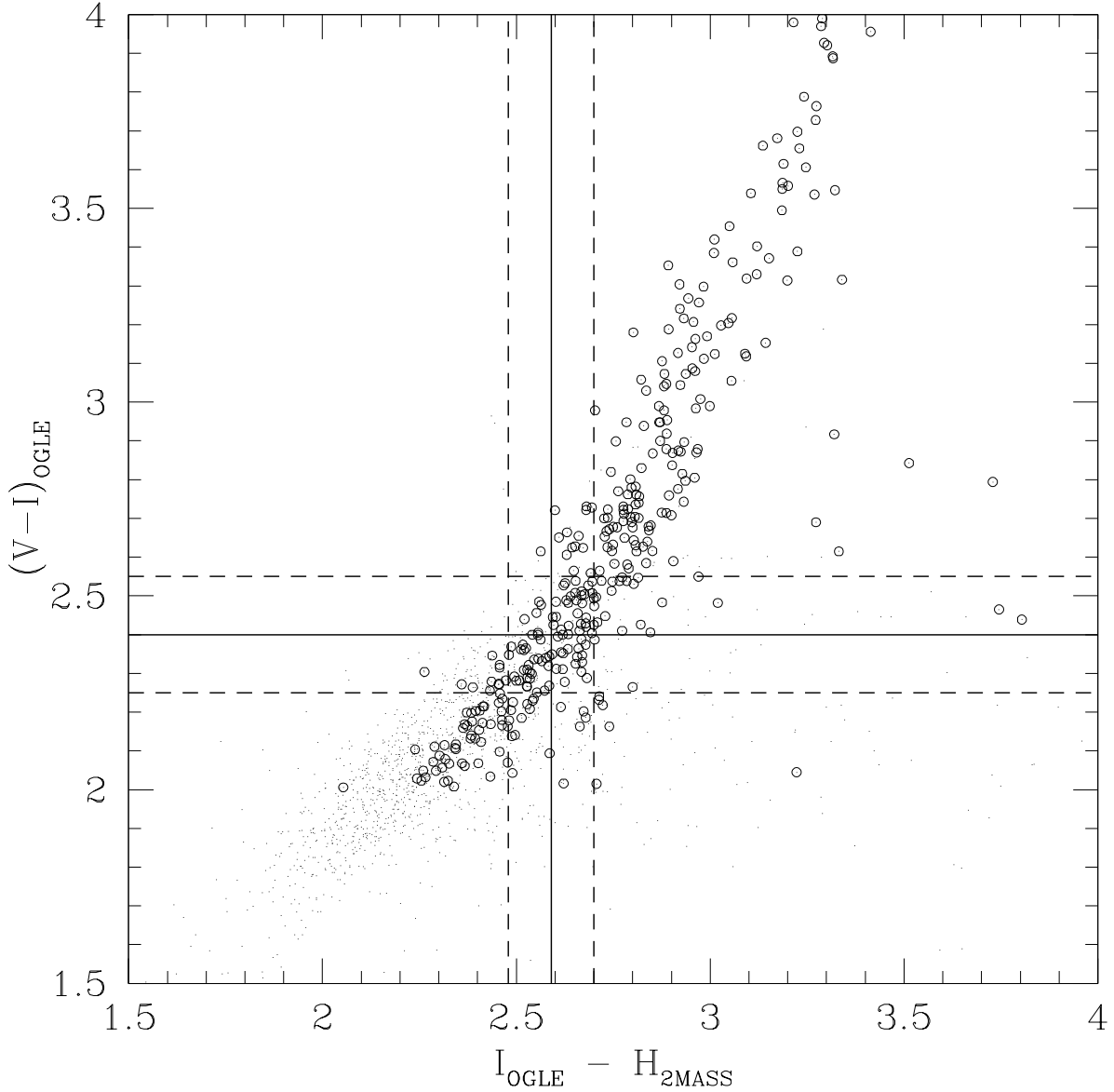}
\caption{
\label{fig:vih}
$(V_{OGLE}-I_{OGLE})/(I_{OGLE}-H_{\rm 2MASS})$ 
color-color diagram. All points are from matching 2MASS $H$-band data
with OGLE-II $V,I$ photometry in a field centered on OGLE-2004-BLG-343.
Stars on the giant branch are shown by open circles. The solid and dashed 
vertical lines represent the source $I_{OGLE}-H_{\rm 2MASS}$ color transformed
from its $I_{OGLE}-H_{\rm \mu{FUN}}$ value and its $1\sigma$ ranges. Their
intersections with the diagonal track of stars give corresponding
$V_{OGLE}-I_{OGLE}$ colors, which are represented by the horizontal lines.
}
\end{figure}

Unfortunately, there are actually very few field stars in the appropriate 
color range. This partly results from the small size ($\sim 2$ arcmin$^2$)
of the $H$-band image and partly from the fact that a large fraction
of stars are either too faint to measure in $V$-band or saturated in
$H$-band. We therefore calibrate the $\mu$FUN $H$-band data by aligning
them to Two Micron All Sky Survery (2MASS) data and generate a $(V-I)/(I-H_{\rm 2MASS})$ color-color diagram
by matching stars from the 2MASS $H$-band data with OGLE-II $V,I$ photometry
in a larger field centered on OGLE-2004-BLG-343. 
We find that $(H_{\rm 2MASS} - H_{\rm \mu{FUN}})=-1.99 \pm 0.01$ from 48 stars 
in common in the field, with a scatter of 0.08 mag. We transform the above
$I-H_{\rm {\mu}FUN}$ color to $I-H_{\rm 2MASS}$ and plot it as a 
{\it vertical line} on a $(V-I)/(I-H_{\rm 2MASS})$ color-color diagram (see Fig.~\ref{fig:vih}). 
From the intersection of the vertical line with the diagonal
track of stars in the field, we infer $V-I = 2.40 \pm 0.15$.

Since the field stars used to make the alignment are giants, this transformation 
would be strictly valid only if the source were a giant as well. 
However, the source star is certainly a dwarf (see Fig.~\ref{fig:fcmd}).
After transforming 2MASS to standard infrared bands \citep{carp01}, we
use the data from Tables II and III of \citet{bb88} to construct
dwarf and giant tracks on a $(V-I)_0/(I-H)_0$ color-color diagram.
  These are approximately coincident for blue stars 
$(I-H)_0<1.6$ but rapidly separate by 0.28 mag in $(V-I)_0$ by $(I-H)_0=1.7$.
In principle, we should just adjust our estimate $(V-I)$ by the 
difference between these two tracks at the dereddened $(I-H)_0$ color
of the source.  Unfortunately, there are two problems with this
seemingly straightforward procedure.  First, the \citet{bb88} giant
track displays a modest deviation from its generally smooth behavior
close to the color of our source, a deviation that is not duplicated
by either the giants in our field or the color-color diagram formed
by combining {\it Hipparcos} and 2MASS data, which both show the
same smooth behavior at this location. Second, if the {\it Hipparcos}/2MASS 
diagram or the \citet{bb88} diagram is reddened using the selective and
total extinctions determined above from the position of the clump,
then the giant tracks do not align with our field giants.  To obtain
alignment, one must use $R_{VI}=2.4$.\footnote{\hei \citet{stutz} found a similar
value using the same method but a different data set. 
However, the $R_{VI}$ we obtained
at the beginning of this section is based on the dereddened magnitude 
of the red clump, which depends on the distance to the Galactic center $R_0$. 
If we were to adopt the new geometric measurement of $R_0 = 7.62\,$kpc \citep{eisenhauer05}, 
rather than the standard value of $R_0 = 8.0\,$kpc, 
we would then have $I_{0,\rm clump} = 14.07$, which would give $R_{VI} = 2.39$.
However this value conflicts still more severely with the 
typical values of $R_{VI} = 1.9$--$2.1$ in bulge fields found by \citet{sumi}.}
The conflict among these three determinations of $R_{VI}$ 
{\hei($1.9$--$2.1$ [\citealt{sumi}], 2.29 [clump], 
$\sim2.4$ [\citealt{stutz}])} is quite a puzzle, but not one that
we can explore here.

The bottom line is that there is considerable uncertainty in the
dwarf-minus-giant adjustment but only in the upward direction.  To
take account of this, we add 0.2 mag error in quadrature to the
upward error bar and finally adopt $V-I=2.4^{+0.25}_{-0.15}$ for
the indirect color determination via the $(I-H)$ measurement.
Finally, we combine this with the direct measurement of $V-I=3.09$
based on the combined $V$ and $I$ light curve.  Because the errors
on the latter measurement are extremely large (and are Gaussian
in flux rather than magnitudes), we determine the probability
distribution for the combined determination numerically
in a flux-based calculation and then convert to magnitudes.  We
finally find
$V-I = (V-I)_{\rm best}\pm\sigma{(V-I)} = 2.60 \pm 0.20$, 
which we show as a magenta point in Figure~\ref{fig:fcmd}. Hence,
\begin{equation}
(V-I)_0 = 1.56 \pm 0.20.
\label{eqn:vi0adopted}
\end{equation}

In contrast to most microlensing events that have been analyzed for
planets, the color of OGLE-2004-BLG-343 is fairly uncertain.  The 
color enters the analysis in two ways.  First, it indicates the
surface brightness and so determines the relation between
dereddened source flux and angular size.  Second, it determines
the limb-darkening coefficient.  

Given the color error, we consider a range of colors in our analysis 
and integrate over this range, just as we integrate over a range
of impact parameters $u_0$ and source sizes (normalized to $\theta_\e$)
$\rho_*$.  We allow colors over the range $2.2<(V-I)<3.0$ corresponding
to $1.16<(V-I)_0<1.96$.  We integrate in steps of 0.1 mag.  For each
color, we adopt a surface brightness such that the source size $\theta_*$
is given by
\begin{equation}
\theta_* = \theta_{(V-I)}\,10^{-0.2(I - I_{\rm best})},
\label{eqn:thetastar}
\end{equation}
where $I$ is the (reddened) apparent magnitude in the model,
$I_{\rm best}=22.24$, and
$\theta_{2.2}\ldots\theta_{3.0} = (0.350,0.371,0.391,0.421,0.466,0.515,0.546,0.580,0.615)\mu$as.
These values are derived from the color/surface-brightness relations for dwarf stars
given in \citet{radius04} using the method as described in \citet{yoo262}. In our actual
calculations, we use the full distribution of source radii, but for reference we note
that the $1 \sigma$ range of this quantity is
\begin{equation}
\theta_* = 0.47\pm0.13\mu{\rm as}.
\label{eqn:thetastar2}
\end{equation}

We find from the models of \citet{claret00} and \citet{haus99} that the 
linear limb-darkening coefficients for dwarfs in our adopted color range
vary by only a few hundredths.  Therefore, for simplicity, we adopt the
mean of these values
\begin{equation}
\Gamma_I = 0.50
\label{eqn:gammai}
\end{equation}
for all colors.  This corresponds to $c=3\Gamma/(2+\Gamma) = 0.60$
in the standard limb-darkening parameterization \citep{afonso00}.

Finally, each model specifies not only a color and magnitude for
the source, but also a source distance.  Evaluation of the likelihood
of each specific combination of these requires a color-magnitude
relation.  We adopt \citep{reid91}
\begin{equation}
M_I = 2.37(V-I)_0 + 2.89
\label{eqn:cmr}
\end{equation}
with a scatter of 0.6 mag.  The ridge of this relation is shown
as a red line segment in Figure~\ref{fig:fcmd}, with the sources 
placed at $10^{-0.15/5}R_0=7.5\,$kpc and reddened according to the red-clump determination, just as was done for the {\it Hipparcos} stars. This track is 
in reasonable agreement with the {\it Hipparcos} stars.

\subsection{Finite-Source Effects}
\label{sec:fs}

\citet{yoo423} define $z_0\equiv u_0/\rho_*$ (where  
$\rho_*=\theta_*/\theta_\e$ is the angular size of 
the source $\theta_*$ in units of $\theta_\e$), which is
a useful parameter to characterize the finite-source effects in single-lens
microlensing events.  We fit the observational data to a set of point-lens models on 
a grid of ($u_0$, $z_0$) and then compare the resulting $\chi^2$ with the best-fit PSPL model.
As mentioned in \S~\ref{sec:cmd}, 
we adopt the limb-darkening formalism of \citet{yoo423} and for simplicity choose
$\Gamma_I=0.50$.

\begin{figure}
\plotone{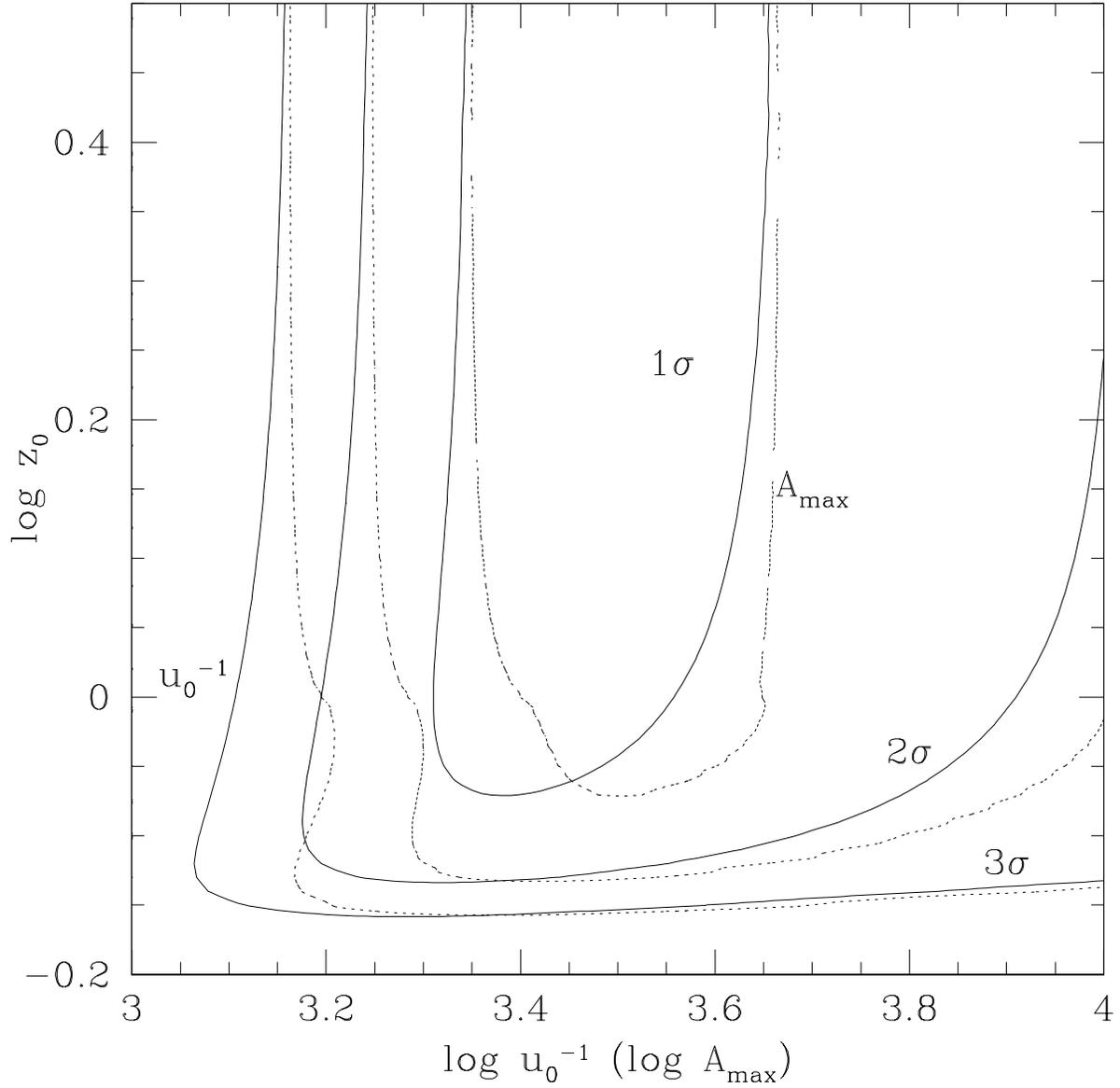}
\caption{
\label{fig:contour1}
Likelihood contours ({\hei$1\sigma, 2\sigma, 3\sigma$}) for finite-source points-lens models
relative to the best-fit PSPL model.
{\hei Contours with $x$-axis as $\log{u_0^{-1}}$ and $\log{A_{\rm max}}$ are 
displayed in solid and dashed lines, respectively. 
}
}
\end{figure}

Figure~\ref{fig:contour1} displays the resulting $\Delta\chi^2$ contours. {\hei It shows
that the 1~$\sigma$ contour extends from $z_0\simeq0.2$ to arbitrarily large  
$z_0$.} This is qualitatively similar to OGLE-2003-BLG-423 as analyzed in \citet{yoo423}. 
However, as we demonstrate in \S~\ref{sec:mcs}, the range of $z_0$ that is consistent
with the Galactic model is quite different for these two events. This is to be expected
since $z_0 = u_0/\rho_* = (u_0/\theta_*)\theta_{\e}$, and $u_0/\theta_*$ is
roughly 8 times smaller for this event, while $\theta_{\e}$ is generally
of the same order.

{\hei Figure~\ref{fig:contour1} 
shows contours for both $A_{\rm max}$ and
$u_0^{-1}$.  For $z_0\equiv u_0/\rho_*>2$, these are very
similar, which is expected because in the absence of finite-source
effects (and for $u_0\ll 1$), $A_{\rm max}=u_0^{-1}$.  Note
that the $A_{\rm max}$ contours are roughly rectangular,
so that while $z_0$ is not well constrained,
the $3\sigma$ lower limit
$A_{\rm max}>10^{3.16}\sim 1450$ is quite well defined.
This shows that although the blending is very severe, it is also
very well constrained, implying that the event's high magnification
is secure.}

\subsection{Monte-Carlo Simulation}
\label{sec:mcs}

We perform a similar Monte-Carlo simulation using a \citet{han96,han03} model 
as described in \citet{yoo423} by taking
into account all combinations of source and lens distances, $D_{\rm l}<D_{\rm s}$, 
uniformly sampled along the line of sight toward the source 
$(l,b)=(4.21,-3.47)$. {\hei The simulation adopts the \citet{gould2000} 
mass function taking into account the bulge main sequence stars, white dwarfs 
(distributed around $0.6 \msun$), neutron stars (narrowly 
peaked at $1.35 \msun$), and stellar-mass black holes. 
This mass function is adequate to describe mass distributions 
of disk lenses except that the disk contains stars
with masses greater than $1\msun$ while the bulge does not. 
However, a disk main sequence star more massive than the Sun will be 
too bright to be the lens star for this event (see Fig.~\ref{fig:fcmd}); 
so for simplicity, we use this mass function for both disk and bulge in
our simulation.} 
In \citet{yoo423}, the source flux is determined from the 
$t_\e$ for each Monte-Carlo event since only the PSPL model is considered at this step. 
However, when finite-source effects are taken into account, 
each $t_\e$ corresponds to a series of $F_{\rm s}$ depending on the source size $\rho_*$,
 so there is no longer a 1-1 correspondence between $F_{\rm s}$ and $t_\e$.
As discussed in \citet{yoo262}, $\theta_*$ can be deduced from the source's dereddened color and
magnitude. Since $\theta_\e$ is known for each simulated event, $\rho_*$ is a direct function of 
$F_{\rm s}$ and the $(V-I)$ color of the source, 
\begin{equation}
\rho_* = {\theta_{(V-I)}\over\theta_\e}\,\sqrt{F_{\rm s}\over F_{\rm best}},
\label{rhostar}
\end{equation}
where $F_{\rm best}$ corresponds to $I_{\rm best}$ in equation~(\ref{eqn:thetastar}).
Using this constraint, we fit the $k$-th Monte-Carlo event to a 
point-lens model with finite-source effects, holding $t_{\e,k}$ fixed at 
the value given by the simulation, for a variety of $(V-I)$ color values inferred 
from \S~\ref{sec:cmd}.
Hence, for the $j$-th $(V-I)$ color and $k$-th Monte-Carlo event,
we have best-fit single-lens light-curve parameters $t_{0,j,k}$, $u_{0,j,k}$, 
$\rho_{*,j,k}$, $F_{{\rm s},j,k}$, $F_{{\rm b},j,k}$, as well as $\Delta\chi^2_{j,k} 
\equiv \chi^2_{j,k} - \chi^2_{\rm PSPL}$. 
We construct a three-dimensional table that includes these six quantities as well as the other
parameters from the Monte-Carlo simulation ($t_{\e,k}$, $\theta_{\e,k}$, 
${D_{{\rm s},k}}$, ${D_{{\rm l},k}}$, ${M_{{\rm l},k}}$, $\Gamma_k$), the Einstein timescale 
and radius, the source and lens distances, the lens mass, and the event rate. From these can 
also be derived two other important quantities, the source absolute magnitude $M_{I,j,k}$ and the
physical Einstein radius $r_{\e,k} \equiv {\theta_{\e,k}\times{D_{{\rm l},k}}}$. This three-dimensional table
is composed of nine two-dimensional tables, one for each $(V-I)_j$ color. Each table 
contains approximately 200,000 rows, one for each simulated event.
To make the notation more compact, we refer to the parameters $a,b,c, ...$ lying in the bin 
$a\in[a_{\min},a_{\max}];b\in[b_{\min},b_{\max}];
c\in[c_{\min},c_{\max}] ...$ as ${\rm bin}(\{a,b,c, ...\})$.  

Similarly to \citet{yoo423}, the posterior probability of $a_i$ lying in 
${\rm bin}(a_i)$ is given by
\begin{equation}
\label{eqn:pai1}
P[{\rm bin}(a_i)] \propto 
\sum_{j,k}
{
{(P_{V-I})_j}
\times(P_{\rm Reid})_{j,k} 
\times\exp[-\Delta\chi_{j,k}^2/2]
\times {\rm BC}[{\rm bin}({\{a_i\}}_{j,k})]}
\times\Gamma_k, 
\end{equation}
where
{\hei$(P_{\rm Reid})_{j,k} = 
\exp(-\{(M_I)_{j,k} - M_{I,\rm Reid}[(V-I)_{0,j}]\}^2
/2[\sigma^2_{\rm Reid}+{(\sigma_{M_I})}^2_{j,k}])$
accounts for the scatter ($\sigma_{\rm Reid}=0.6$) in $M_I$ 
about the Reid relation plus the dispersion ${(\sigma_{M_I})}_{j,k}$ 
from light-curve fitting, 
$(P_{V-I})_j = \exp\{-[(V-I)_j-(V-I)_{\rm best}]^2/2\sigma^2_{(V-I)}\}$ reflects the 
 uncertainty in $V-I$ color, and ${\rm BC}$ is a boxcar function defined by
${\rm BC}[{\rm bin}(a)] \equiv \Theta\left(a - a_{\min}\right)
\times\Theta\left(a_{\max} - a\right)$.}

\begin{figure}
\includegraphics[width = 480pt, height = 420pt]{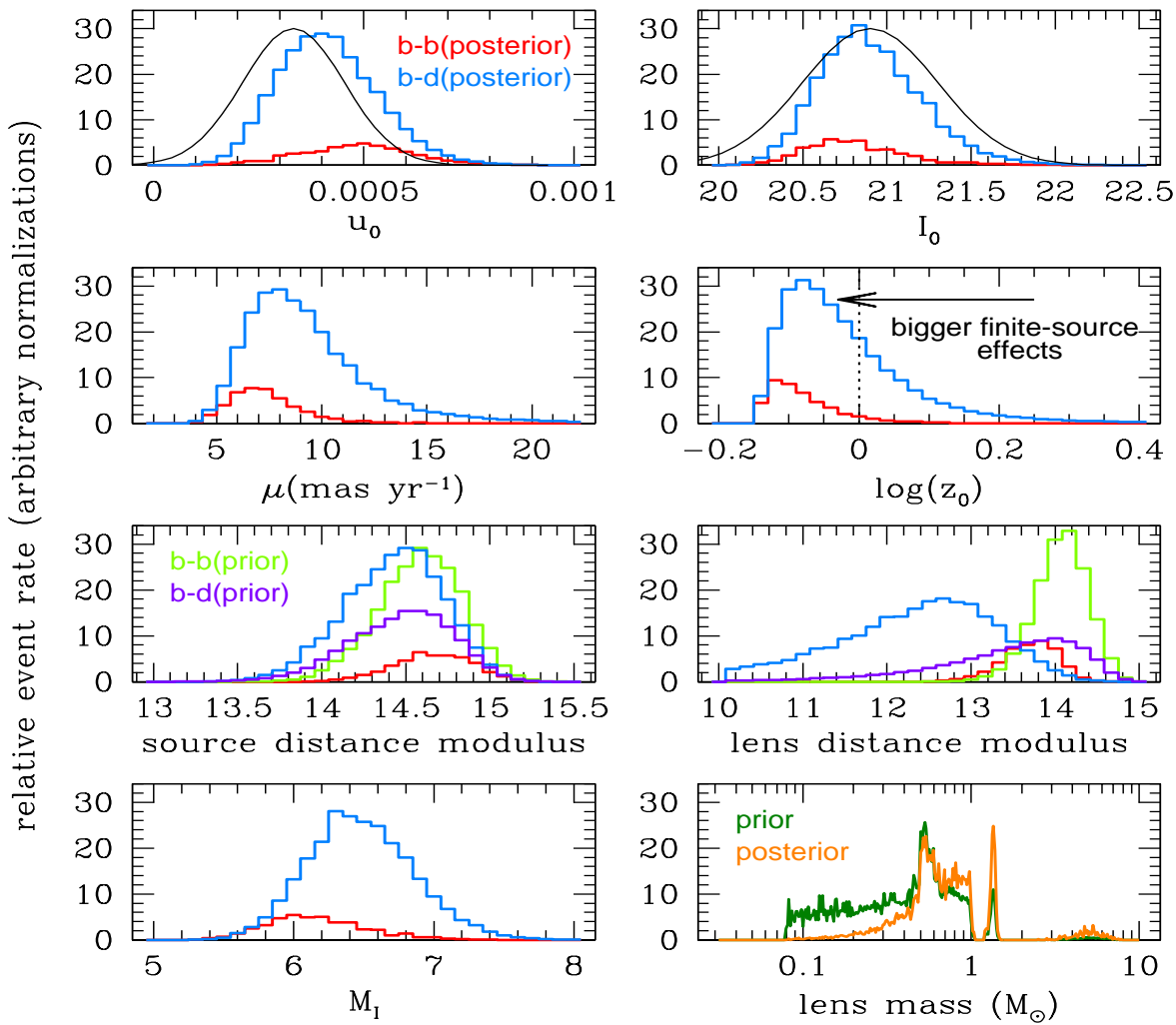}
\caption{
\label{fig:fmcs}
{\hei Probability distributions of $u_0$, dereddened apparent $I$-band magnitude
of the source $I_0$, proper motion $\mu$, $\log(z_0)$, source distance modulus,
lens distance modulus, absolute $I$-band magnitude of the source $M_I$, and lens mass
for Monte Carlo events toward the line of sight of OGLE-2004-BLG-343.} 
Blue histograms represent the posterior probability distributions for
bulge-disk microlensing events while 
red ones represent the posterior probability distributions for 
bulge-bulge events. In the source and lens 
distance-modulus panels, histograms in purple and green represent
the prior probability distributions for bulge-disk and bulge-bulge events, respectively. 
The black Gaussian curves in the $u_0$ and $I_0$ panels show probability
 distributions from PSPL light-curve fitting alone. In the lens mass panel, 
the dark green histogram shows the prior probability distribution, while
the orange histogram represents the posterior distribution.}
\end{figure}

Figure~\ref{fig:fmcs} shows the posterior probability distributions of 
various parameters, including $u_0$, dereddened apparent $I$-band magnitude 
of the source $I_0$, proper motion $\mu$, $z_0$, source distance modulus, 
lens distance modulus, absolute $I$-band magnitude of the source $M_I$, and lens mass. 
The blue and red histograms represent bulge-disk events and 
bulge-bulge events, respectively. 
{\hei The relative event rate is normalized in the same way for both 
bulge-disk and bulge-bulge events. The total rate for bulge-disk events 
is about $6$ times larger than that for bulge-bulge events, which
means that the Monte-Carlo simulation tends strongly to favor bulge-disk events. 

The Einstein radii are on average smaller for bulge-bulge events than for bulge-disk events, 
and as a result, the bulge-bulge events tend to have bigger $\rho_*$ and hence smaller 
$z_0$. However, the top right panel of Figure~\ref{fig:fmcs} shows that the
$z_0$ probability distributions have similar shapes for both bulge-bulge
 and bulge-disk events. This is because
the (lack of) finite-source effects constrain $z_0 \ga 0.7$ at the $3 \sigma$ level 
(see Fig.~\ref{fig:contour1}), which cuts off the lower end of the $z_0$ distributions
 for both categories of events. Since bulge-disk events have smaller $\rho_*$ than 
bulge-bulge 
events, the $u_0$ posterior probability distribution peaks at a lower value for the
former. Furthermore, since $z_0 \ga 0.7$, the proper 
motion is constrained to be $\mu = {\theta_*z_0 / t_{\eff}} \ga 7{\rm mas/yr}$, 
which is typical of bulge-disk events but {\hei $\ga 2$ times} 
the proper motion of typical bulge-bulge events. }

Figure~\ref{fig:fmcs} also shows the 
distributions of $u_0$ and $I_0$ from the light-curve data alone by a black solid line.
 In strong contrast to the corresponding diagrams for OGLE-2003-BLG-423 presented 
by \citet{yoo423}, the light-curve based parameters agree quite well with
the Monte-Carlo predictions. In the source distance-modulus panel,
 the prior distributions for bulge-disk and bulge-bulge events are shown in purple 
and green histograms, respectively. Again, distinct from OGLE-2003-BLG-423, the most 
likely source distances of this event agree reasonably well with typical values 
from the prior distributions. Moreover, from Figure~\ref{fig:fmcs}, the peak values 
of source $M_I$ distributions are also in good agreement with those derived from the Reid relation
{\hei ($M_I = 6.59$ for $\left[V-I\right]_0 = 1.56$, see eq.~[\ref{eqn:cmr}])}. 
Therefore, the source of this event 
shows very typical characteristics as represented by the Monte-Carlo simulation.
{\hei Also unlike OGLE-2003-BLG-423, the probability that $z_0 \la 1$ is very high
for both bulge-disk and bulge-bulge events. Therefore, finite-source effects 
must be taken into account in the analysis of this event.}
In addition to the posterior probability distribution ({\it orange}) of the lens mass,
 the prior distribution ({\it dark green}) is displayed in Figure~\ref{fig:fmcs} as well. 
In microlensing analyses, 
the lens mass is commonly assigned a ``typical'' value (for example, $0.3\msun$).
 However, Figure~\ref{fig:fmcs} shows that, lenses with relatively high mass are strongly 
favored for this event as compared to the prior distribution. Detailed discussions 
on the lens properties are presented in \S~\ref{sec:luminouslens}.

\section{Detecting Planets}
\label{sec:dp}

While there are no obvious deviations from point-lens behavior in the
light curve of OGLE-2004-BLG-343 at our adopted threshold of
$\Delta\chi^2_{\rm min}=60$, planetary deviations might be difficult
to recognize by eye.  We must therefore conduct a systematic search
for such deviations.  Logically, this search should precede the
second step of testing to determine what planets we could have
detected had they been there.  However, as a practical matter
it makes more sense to first determine the range of parameter
space for which we are sensitive to planets because it is only
this range that needs to be searched for planets. We therefore
begin with this detection efficiency calculation.

\subsection{Detection Efficiency}
\label{sec:de}

{\hei As reviewed in \S~\ref{sec:subintro}},
a variety of methods have been proposed to calculate the planetary sensitivities of 
microlensing events, either in predicting planetary detection efficiencies theoretically 
or in analyzing real observational data sets. In those methods, $\Delta{\chi^2}$ is often 
calculated by subtracting the $\chi^2$ of
single-lens models from that of the binary-lens models to evaluate detection sensitivities. 
However, the ways in which single-lens and binary-lens models are compared differ from study 
to study. 
As noted by \citet{griest98} and \citet{gaudi00}, for real planetary light curves, 
the lens parameters are not known {\it a priori}. Therefore, $\Delta{\chi^2}$ will generally be
exaggerated if one subtracts {\hei from} the binary lens model the single-lens model 
that has the same $t_0$, $u_0$, and $t_\e$ instead of the best-fit single-lens model 
to the binary light curve. One important factor contributing to this exaggeration is that the 
center of the magnification pattern (referred to as the center of the 
caustic in \citealt{yoo423}) in the binary-lens case is no longer the position of 
the primary star as it is in the single-lens model \citep{dominik99,an02}. 
Therefore light-curve parameters such as $u_0$ and $t_0$ will shift correspondingly. 
We find that by not taking into account this effect and directly comparing the 
simulated binary (i.e., planetary) light curve with the best-fit single-lens model 
to the data, \citet{abe04} exaggerate the planetary sensitivity of MOA 2003-BLG-32/OGLE 
2003-BLG-219, although it remains the most sensitive event analyzed to 
date (see Appendix~\ref{sec:moa}).

Following \citet{gaudi00}, planetary systems are characterized by a planet-star mass 
ratio $q$, planet-star separation in Einstein radius $b$, and the angle $\alpha$ of 
the source trajectory relative to the planet-star axis. In our binary-lens calculations,
$u_0$ and $t_0$ are defined with respect to the center of magnification discussed above.
According to \citet{gaudi00}, the next step is to fit the data to both PSPL 
models and binary-lens models with a variety of $(b,q,\alpha)$ and calculate 
$\Delta{\chi^2{(b,q,\alpha)}} = \chi^2{(b,q,\alpha)} - \chi^2_{\rm PSPL}$. 
Then $\Delta{\chi^2{(b,q,\alpha)}}$ is compared with a threshold value $\chi^2_{\rm thres}$: 
if $\Delta{\chi^2} > \chi^2_{\rm thres}$ then a planet with parameters $b$, $q$, and $\alpha$ 
is claimed to be excluded while it is detected if $\Delta{\chi^2} < -\chi^2_{\rm thres}$. 
The $(b,q)$ detection efficiency is then obtained by integrating 
$\Theta(\Delta{\chi^2}-\Delta{\chi_{\rm thres}^2})$ over $\alpha$ in the 
exclusion region at fixed $(b,q)$, where $\Theta$ is a step function. 
However, \citet{gaudi02} point out that for events with poorly constrained light-curve parameters, 
which is the case for OGLE-2004-BLG-343, this method will significantly underestimate the 
sensitivities since the binary-lens models will minimize the $\chi^2$ over the relatively
large available parameter space. 
As discussed in \citet{yoo423}, the detection efficiency should be 
evaluated at a series of allowed $u_0$ values. To take finite-source effects into account, 
we generate a grid of permitted $(u_0,\rho_*)$, and each $(u_{0,m},\rho_{*,m})$ bin is 
associated with the probability $P[{\rm bin}(\{u_{0,m},\,\rho_{*,m}\})]$ obtained 
using the following equation:

\begin{equation}
P[{\rm bin}(\{u_{0,m},\,\rho_{*,m}\})] \propto \sum_{j,k}{P_{m,j,k}}
\label{eqn:pai2}
\end{equation}
where
\begin{equation}
{
P_{m,j,k} =
{(P_{V-I})_j}
\times
(P_{\rm Reid})_{j,k}
\times
\exp[-\Delta\chi_{j,k}^2/2]
\times
{\rm BC}[{\rm bin}(\{u_{0,m}\}_{j,k})]\times{\rm BC}[{\rm bin}(\{\rho_{*,m}\}_{j,k})]
}
\times
\Gamma_k
\label{eqn:pai3}
\end{equation}

If the light-curve parameters were well-constrained, the approaches
of \citet{gaudi00} and \citet{rhie00} would be
very nearly equivalent, with the former retaining a modest
philosophical advantage, since it uses only the observed light curve and
does not require construction of light curves for hypothetical
events.  However, because in our case these parameters are not
well constrained, the \citet{gaudi00}  approach would
require integration over all binary-lens parameters (except $F_{\rm s}$
and $F_{\rm b}$).  Regardless of its possible philosophical advantages,
this approach is therefore computationally prohibitive in the
present case.  
We therefore do not follow \citet{gaudi00}, 
but instead construct a binary light curve with the same 
observational time sequence and 
photometric errors as the OGLE observations of OGLE-2004-BLG-343, for 
each $(b,q,\alpha\,;u_0,\rho_*)$ combination and the 
associated probability-weighted parameters $\bdv{a}_{{\rm lc}}$: 
$t_0$, $t_\e$, $F_{\rm s}$ and $F_{\rm b}$ in the $m$-th $(u_0,\rho_*)$ bin,
\begin{equation}
{\bdv{a}_{\rm lc(weighted),m}} =
{
{
\sum_{j,k}\limits
{P_{m,j,k}{{\bdv{a}_{{\rm lc},j,k}}}
}
\over{\sum_{{j,k}}\limits{P_{m,j,k}}}
}
}.
\label{eqn:aprob}
\end{equation} 

Then each simulated binary light curve $(b,q,\alpha\,;u_{0,m},\rho_{*,m})$
is fitted to a single-lens model with finite-source effects 
whose best fit yields $\chi^2(b,q,\alpha\,;u_{0,m},\rho_{*,m})$.
Another set of artificial binary light curves is generated 
under the assumption that OGLE had triggered a dense 
series of observations following the internal alert at HJD$'$ 3175.54508.
These cover the peak of the event with the normal OGLE frequency and are used to compare 
results with those obtained from the real observations.

Magnification calculations for a binary lens with finite-source effects are very
time-consuming. 
Besides $(b,q,\alpha)$, our calculations are also performed on $(u_0,\rho_*)$ grids, 
two more dimensions 
than in any previous search of a grid of models with finite-source effects 
included. This makes our computations
extremely expensive, comparable to those of \citet{gaudi02}, which equaled 
several years of processor time. Therefore, we have developed two new 
binary-lens finite-source 
algorithms to perform the calculations, as discussed in detail in Appendix~\ref{sec:code}.

In principle, we should consider the full range of $b$, i.e., $0<b<\infty$; in practice,
it is not necessary to directly simulate $b<1$ due to the famous $b\leftrightarrow  b^{-1}$ 
degeneracy \citep{dominik99a,jin05}. 
Instead, we just map the $b>1$ results onto $b<1$ except for the isolated sensitive zones along 
the $x-$axis caused by planetary caustics perturbations.

We define the planetary detection efficiency $\epsilon(b,q)$ as the probability that 
an event with the same characteristics as  OGLE-2004-BLG-343,
 except that the lens is a planetary system with configuration of $(b,q)$, is 
inconsistent with the single-lens model (and hence would have been detected),

\begin{eqnarray}
\epsilon(b,q,\alpha) &=&
\{\sum_{m}\limits\Theta \left[\chi^2(b,q,\alpha\,;u_{0,m},\rho_{*,m}) -\Delta\chi^2_{\rm thres}\right] \nonumber \\
&&\times P [{\rm bin}(\{u_{0,m},\,\rho_{*,m}\})]\} \nonumber \\
&&\times \{\sum_{m}\limits P[{\rm bin}(\{u_{0,m},\,\rho_{*,m}\})]
\}^{-1}
\label{eqn:epsdef1}
\end{eqnarray}
and
\begin{equation}
\epsilon(b,q)={1\over2\pi}\int_0^{2\pi}{\epsilon(b,q,\alpha)}d\alpha.
\label{eqn:epsdef2}
\end{equation}

\subsection{Constraints on Planets}
\label{sec:res}

\begin{figure}
\plotone{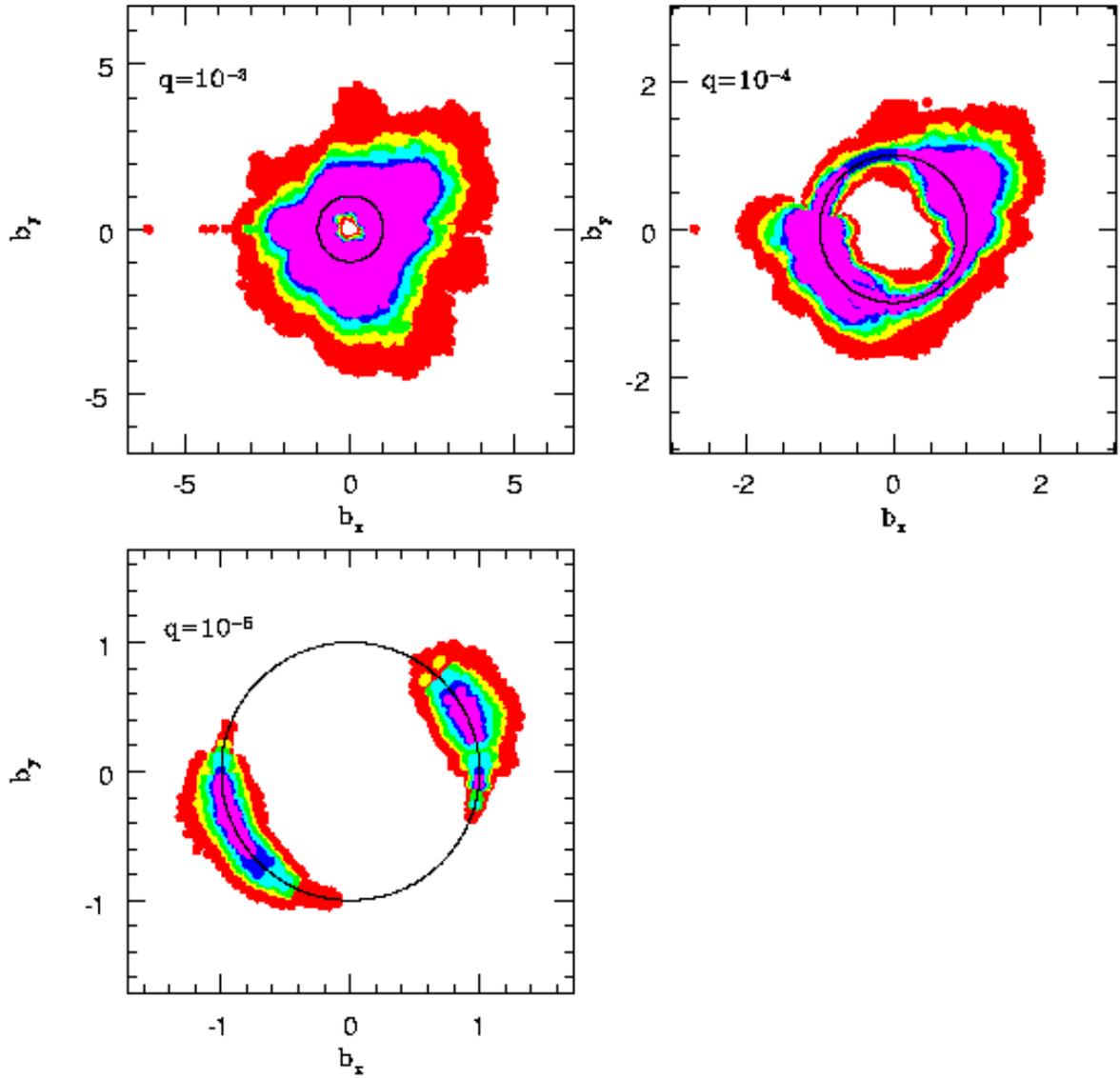}
\caption{
\label{fig:realcircs}
(For real data) Planetary detection efficiency for mass ratios $q=10^{-3}$, $10^{-4}$, 
and $10^{-5}$ for OGLE-2004-BLG-343 as a function of the planet-star separation 
$b_x = b\times{\cos{\alpha}}$ and $b_y = b\times{\sin{\alpha}}$ in 
the units of ${\theta}_\e$ where $\alpha$ is the angle of planet-star axis relative 
to the source-lens direction of motion.
Different colors indicate 10\% ({\it red}), 25\% ({\it yellow}), 50\% ({\it green}), 
75\% ({\it cyan}), 90\% ({\it blue}) and 100\% ({\it magenta}) efficiency.
The black circle is the Einstein ring, i.e., $b=1$.
}
\end{figure}

Figure~\ref{fig:realcircs} shows the planetary detection efficiency
of OGLE-2004-BLG-343 for
planets with mass ratios $q=10^{-3}$, $10^{-4}$, and $10^{-5}$, as
a function of $b$, the planet-star separation (normalized to $\theta_\e$),
and $\alpha$, the angle that the moving source makes with the binary axis 
passing the primary lens star on its left.
Different colors indicate 10\%, 25\%, 50\%, 75\%, 90\% and 100\% efficiency.
Note that the contours are elongated
along an axis that is roughly $60^\circ$ from the vertical (i.e.,
the direction of the impact parameter for $\alpha=0$).  This reflects the fact that
the point closest to the peak occurs at $t=2453175.77626$ when
$(t-t_0)/t_\e = 2.16u_0$, and so when the source-lens separation is
at an angle $\tan^{-1} 2.16 = 65^\circ$.
For $q=10^{-3}$, the region of 100\% efficiency extends 
through $360^\circ$ within about one octave on either side of 
the Einstein ring.  However, at lower mass ratios there 
is 100\% efficiency only in restricted areas
close to the Einstein ring and along the above-mentioned principal axis.

\begin{figure}
\plotone{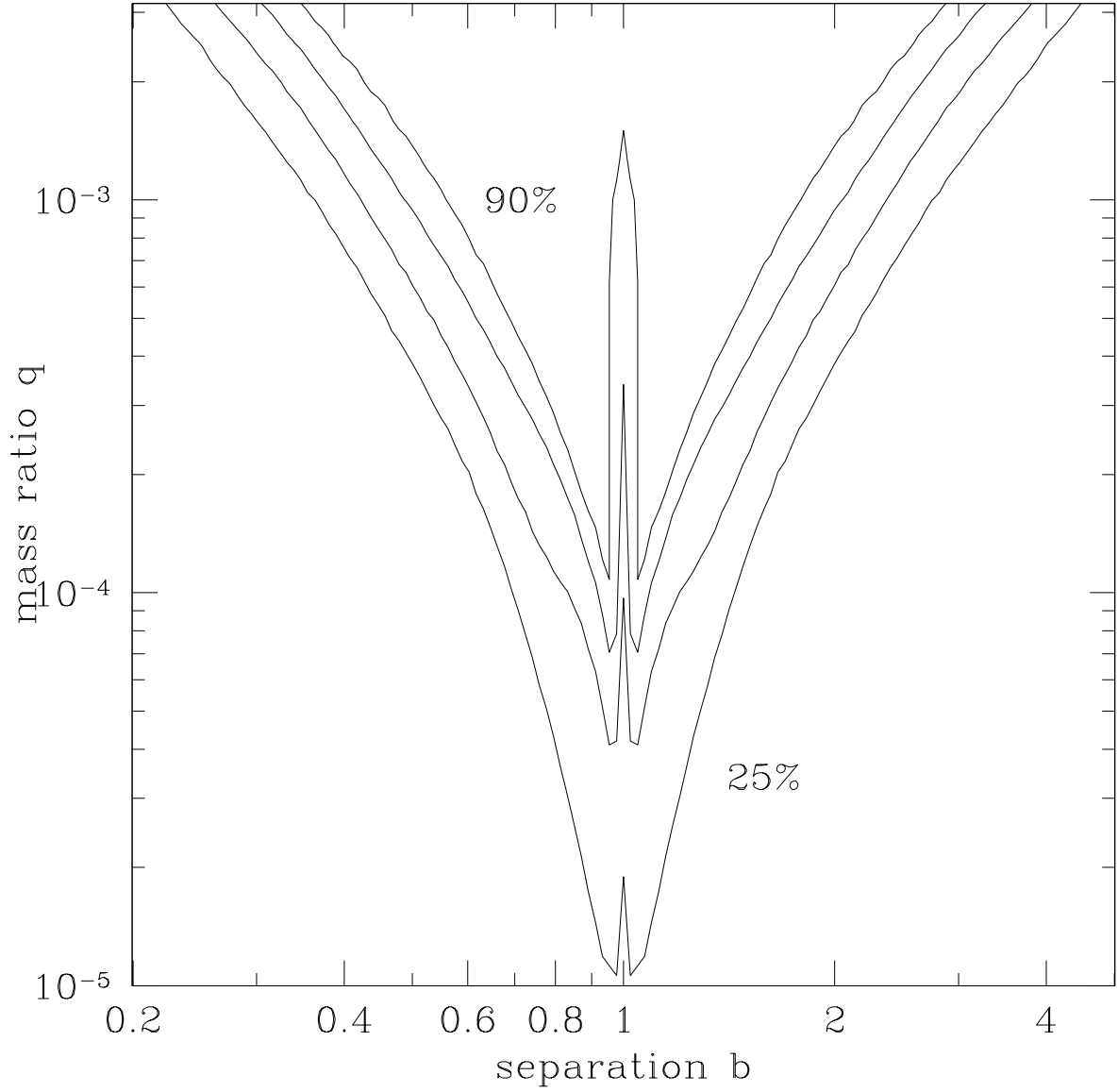}
\caption{
\label{fig:realsum}
(For real data) Planet detection efficiency of OGLE-2004-BLG-343 as a function of the 
planet-star separation $b$ (in the units of $r_\e$) and planet-star
mass ratio $q$. The contours indicate 25\%,  50\%, 75\%, and 90\%
efficiency.
}
\end{figure}

Figure~\ref{fig:realsum} summarizes an ensemble of all figures similar to
Figure~\ref{fig:realcircs},
but with $q$ ranging from $10^{-2.5}$ to $10^{-5.0}$ in
0.1 increments.  To place this summary in a single figure, we
integrate over all angles $\alpha$ at fixed $b$.  Comparison of this figure to
Figure 8 from \citet{gaudi02} shows that the detection efficiency of OGLE-2004-BLG-343
is similar to that of MACHO-1998-BLG-35 and OGLE-1999-BUL-35 despite the
fact that their maximum magnifications are $A_{\rm max}\sim 100$, roughly
30 times lower than OGLE-2004-BLG-343.  Of course, part of the reason
is that OGLE-2004-BLG-343 did not actually probe as close as 
$u=u_0\sim 1/3000$ because no observations were taken near the peak.  However,
observations {\it were} made at $u\sim 1/1200$, about 12 times closer
than in either of the two events analyzed by \citet{gaudi02}.  One
problem is that because the peak was not well covered, there are planet
locations that do not give rise to observed perturbations at all.
But this fact only accounts for the anisotropies seen in
Figure~\ref{fig:realcircs}. More fundamentally, even perturbations that
do occur in the regions that are sampled by the data
can often be fitted to a point-lens light curve by ``adjusting'' the
portions of the light curve that are not sampled.

Note the central ``spike'' of reduced detection efficiency plots near $b=1$.
As first pointed out by \citet{bennettrhie96}, this is due to the extreme weakness of the
caustic for nearly resonant ($b\sim1$) small mass-ratio ($q\ll1$) binary lenses.

\subsection{No Planet Detected
\label{sec:noplanet}}

Based on the detection efficiency levels we obtained in \S~\ref{sec:res}, 
we fit the observational data to binary-lens models
 to search for a planetary signal in the regions with 
efficiency greater than zero from $q = 10^{-5}$ to $q = 10^{-2.5}$. 
We find no binary-lens
models satisfying our detection criteria. In fact, the total $\chi^2$ contributions to
the best-fit single-lens model of the observational points over the peak (
HJD $= 2453175.5 - 2453176.0$) are no more than 30, so even if all of these deviations 
were due to a planetary perturbation, such a binary-lens solution would not
easily satisfy our $\Delta{\chi^2}=60$ detection criteria.
Therefore there are no planet
detections in OGLE-2004-BLG-343 data.

\subsection{Fake Data
\label{sec:resfake}}

Partly to explore further the issue of imperfect coverage of the peak, 
and partly to understand how
well present microlensing experiments can probe for planets, we
now ask what would have been the detection efficiency of OGLE-2004-BLG-343
if the internal alert issued on HJD$'$ 3175.54508 had been acted upon.

Of course, since the peak was not covered, we do not know exactly what
$u_0$ and $\rho$ for this event are. However, for purposes of this
exercise, we assume that they are near the best fit as determined from 
a combination of the light-curve fitting and the Galactic Monte Carlo, and
for simplicity, we choose $u_0= 0.00040, \rho_*=0.00040$ which is very close
to the best-fit combination. We then form a fake light curve sampled at
intervals of 4.3 minutes, starting from the alert and continuing to
the end of the actual observations that night. 
{\hei This sampling reflects the intense rate of OGLE follow-up observations
actually achieved during this event (see \S~\ref{sec:data}).}
 We assume errors similar
to those of the actual OGLE data at similar magnifications. 
For those points that are brighter than the brightest OGLE point, the minimum 
actual photometric errors are assigned.
We also assume that the color information is known exactly in this case
to be $V-I=2.6$. We then analyze these fake data in exactly the same way 
that we analyze the real data.  In contrast to the real data, however, 
we do not find a finite range of $z_0\equiv u_0/\rho_*$ that are consistent with 
the fake data. Rather, we find that all consistent parameter combinations 
have $z_0=1$ almost identically.  We therefore consider only a 
one-dimensional set of $(u_0,\rho_*)$ combinations subject to this constraint.

\begin{figure}
\plotone{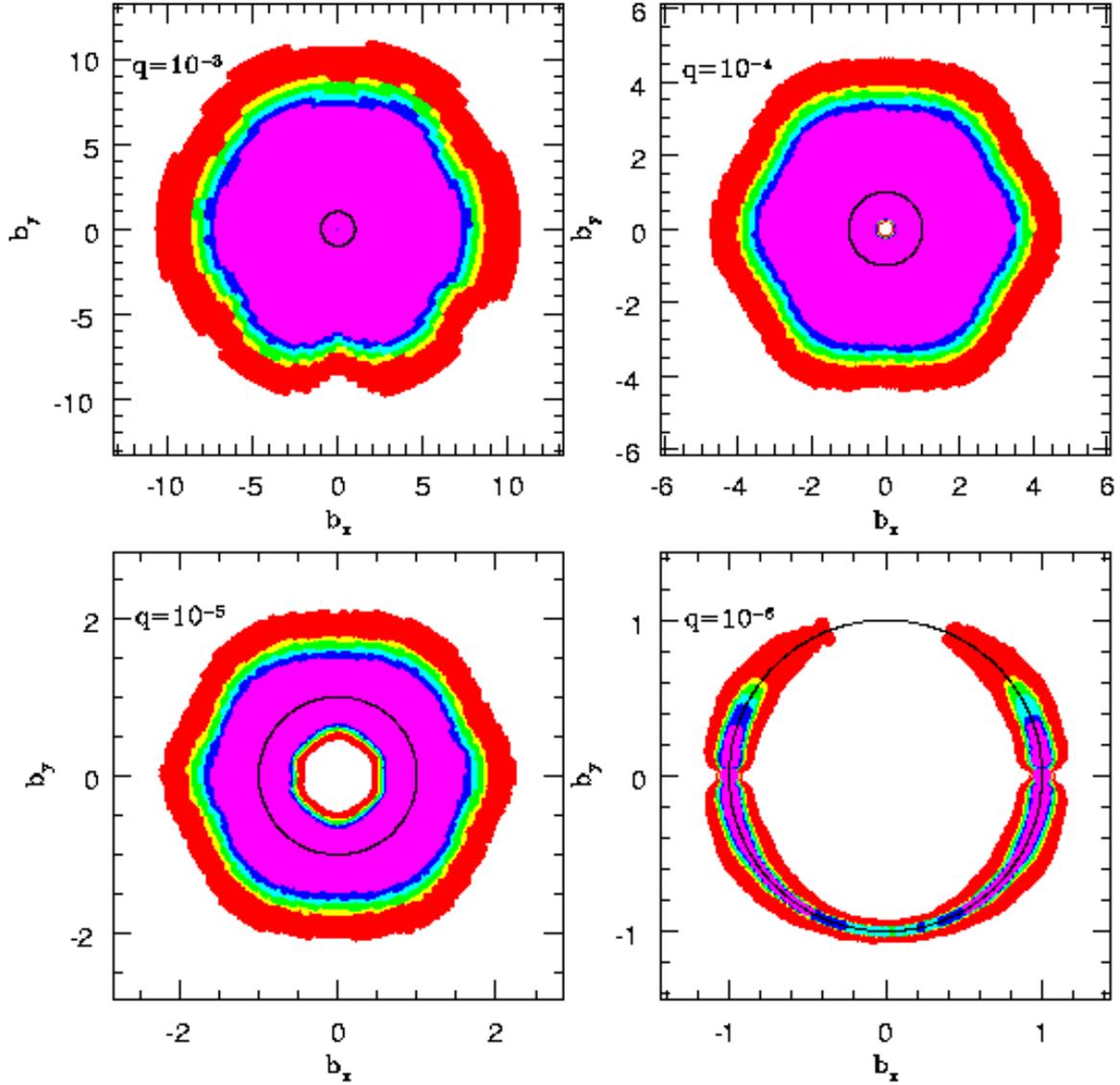}
\caption{
\label{fig:fakecircs}
(For fake data) Planetary detection efficiency for mass ratios 
$q=10^{-3}$, $10^{-4}$, $10^{-5}$ and 
$10^{-6}$ of OGLE-2004-BLG-343 augmented by simulated data points 
covering the peak as a function of the planet-star separation $b_x$ and $b_y$ in
the units of ${\theta}_\e$. 
Different colors indicate 10\% ({\it red}), 25\% ({\it yellow}), 50\% ({\it green}), 
75\% ({\it cyan}), 90\% ({\it blue}) and 100\% ({\it magenta}) efficiency.
The black circle is the Einstein ring, i.e., $b=1$.
}
\end{figure}

Figure~\ref{fig:fakecircs} is analogous to Figure~\ref{fig:realcircs}
except that the panels show planet sensitivities for
$q=10^{-3}$, $10^{-4}$, $10^{-5}$, and $10^{-6}$, that is, an extra
decade.  In sharp contrast to the real data, these sensitivities are
basically symmetric in $\alpha$, except for the lowest value of $q$.
Sensitivities at all mass ratios are dramatically improved.  For example,
at $q=10^{-3}$, there is 100\% detection efficiency over 1.7 dex in $b$
($1/7 \la b \la 7$).  Even at $q=10^{-5}$ (corresponding to an Earth-mass
planet around an M star), there is 100\% efficiency over an octave
about the Einstein radius.

\begin{figure}
\plotone{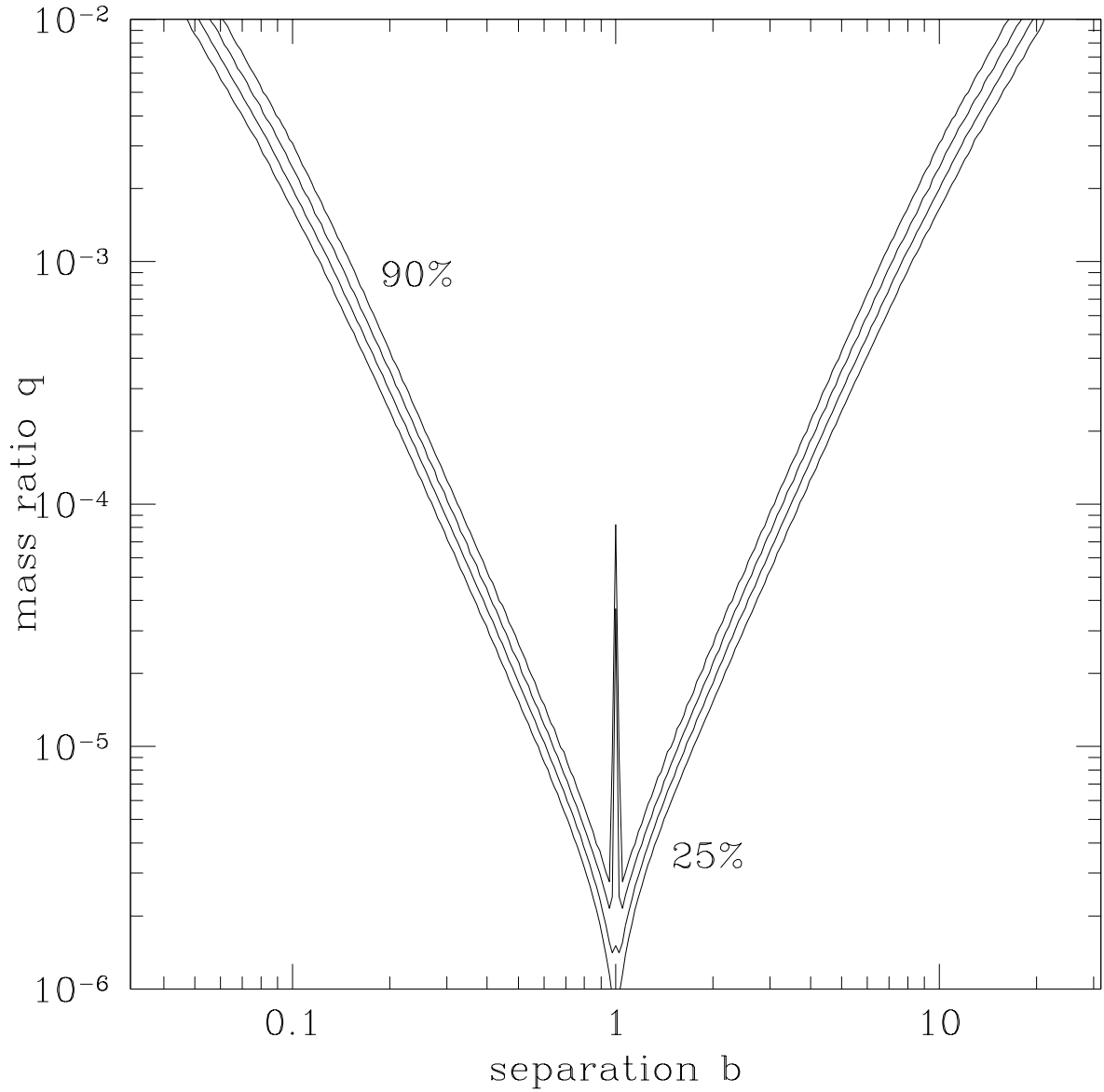}
\caption{
\label{fig:fakesum}
(For fake data) Planetary detection efficiency of OGLE-2004-BLG-343 
augmented by simulated data 
points covering the peak. as a function of the planet-star separation $b$ 
(in the units of ${\theta}_\e$) and planet-star mass ratio $q$. The contours 
represent 25\%,  50\%, 75\%, and 90\% efficiency.
}
\end{figure}

Figure~\ref{fig:fakesum} is the fake-data analog of Figure~\ref{fig:realsum}.
It shows that this event would have been sensitive to extremely {\hei low mass-ratios,}
 lower than those accessible to any other technique other than
pulsar timing.

\subsection{Detection Efficiency in Physical Parameter Space}
\label{sec:realspe}

\begin{figure}
\plotone{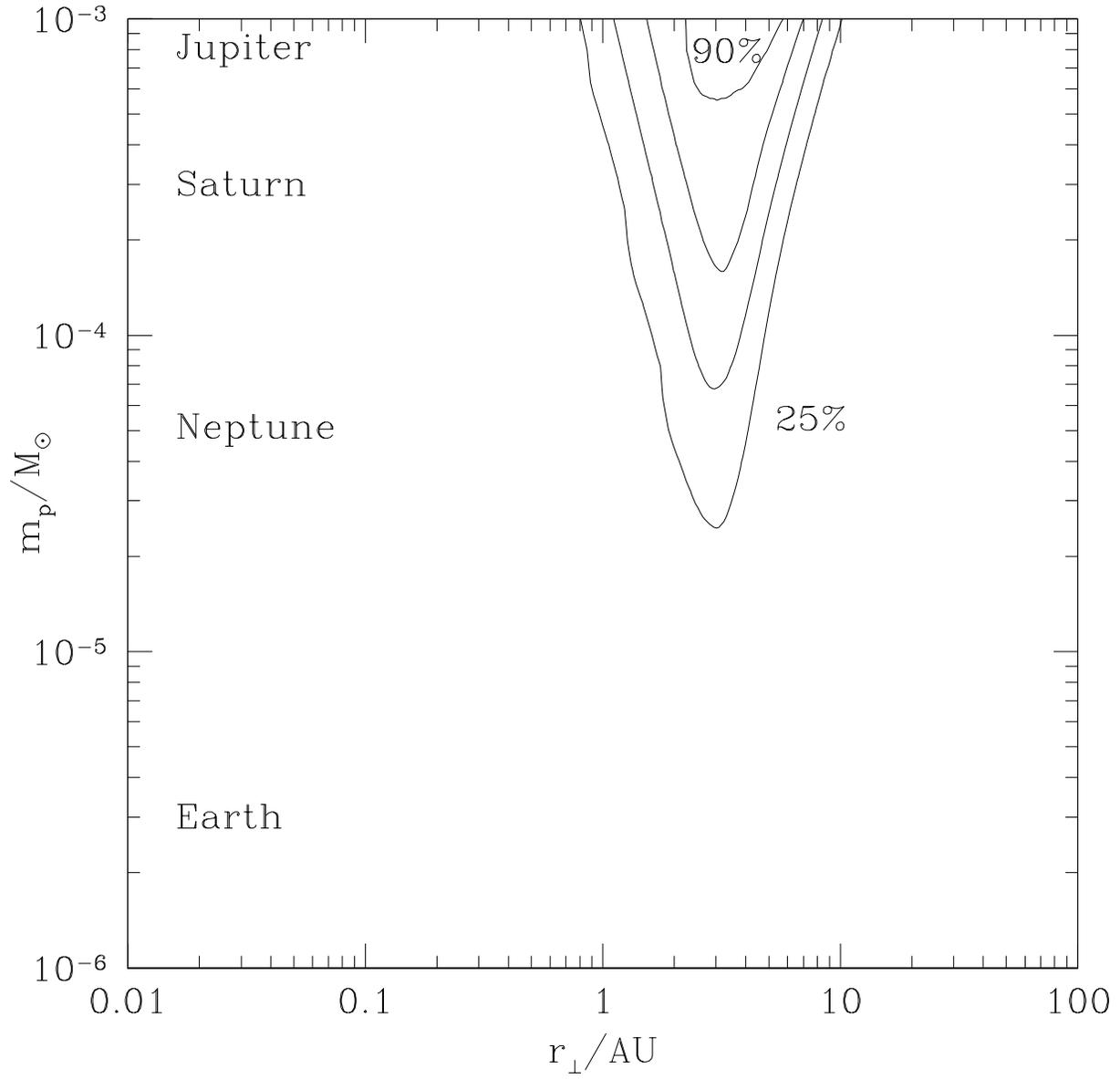}
\caption{
\label{fig:mrreal}
(For real data) Planetary detection efficiency as a function of 
$r_{\perp}$, the physical projected star-planet distance and $m_p$, the planetary mass for 
OGLE-2004-BLG-343. The contours represent 25\%, 50\%, 75\%, and 90\% efficiency.}
\end{figure}

\begin{figure}
\plotone{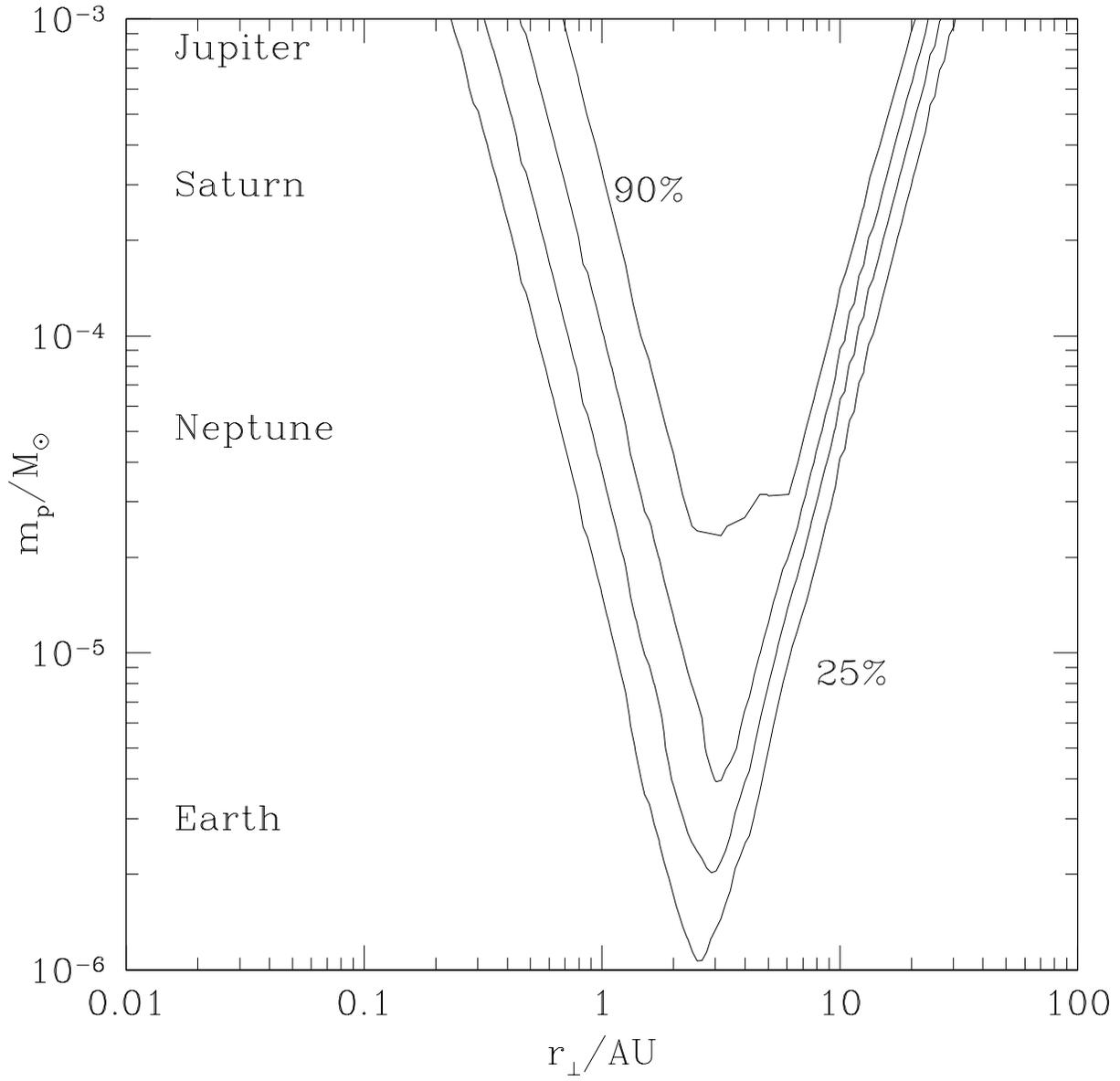}
\caption{
\label{fig:mrfake}
(For fake data) Planetary detection efficiency as a function of  $r_{\perp}$, the physical 
projected star-planet distance and $m_p$, the planetary mass for 
OGLE-2004-BLG-343 augmented by simulated data points covering the peak. 
The contours represent 25\%, 50\%, 75\%, and 90\% efficiency.
}
\end{figure}

One of the advantages of the Monte Carlo approach of \citet{yoo423} is
that it permits one to evaluate the planetary detection efficiency in the
space of the {\it physical} parameters, planet mass and
projected physical separation ($m_p,r_\perp$), rather than just the 
microlensing parameters $(b,q)$.  Figures~\ref{fig:mrreal} and~\ref{fig:mrfake} 
show this detection efficiency for the real and fake data, 
respectively. The fraction of Jupiter-mass planets that could have been
detected from the actual data stream is greater than 
$25\%$ for $0.8{\rm AU}\la{r_{\perp}}\la{10{\rm AU}}$ and is greater 
than $90\%$ for $2{\rm AU}\la{r_{\perp}}\la6{\rm AU}$. There is also 
marginal sensitivity to Neptune-mass planets.
However, the detection efficiencies would have been significantly 
enhanced had the FWHM around the peak been observed, {\hei as previously
discussed by \citet{moa02}}. 
For the fake data, more than $90\%$ of Jupiter-mass planets in the range
$0.7{\rm AU}\la{r_{\perp}}\la20{\rm AU}$ and more than $25\%$ with
$0.3{\rm AU}\la{r_{\perp}}\la30{\rm AU}$ would have been detected. Indeed,
some sensitivity would have extended all the way down to Earth-mass planets. 

\section{Luminous Lens?
\label{sec:luminouslens}}

Understanding the physical properties of their host stars
is a major component of the study of extra-solar planets.
It is especially important to know the mass and distance
of the lens star for planets detected by microlensing
because only then can we accurately determine the planet's
mass and physical separation from the star.  Obtaining
similar information for microlensing events that are
unsuccessfully searched for planets enables more
precise estimates of the detection efficiency.  There
are only two known ways to determine the mass and distance
of the lens: either measure both the microlensing parallax and
the angular Einstein radius (which are today possible for only
a small subset of events) or directly image the lens.
In most cases the lens is either entirely invisible
or is lost in the much brighter light of the source.

A simple argument suggests, however, that in extremely
high-magnification events like OGLE-2004-BLG-343, the
lens will often be easily visible and, indeed, it
is the lens that is unknowingly being monitored, with
the source revealing itself only in the course of the
event.  Events of magnification $A_\max$ require that
the source be much smaller than the Einstein radius,
$\theta_*\la 2\theta_\e/A_\max$. Since
$\theta_\e=\sqrt{\kappa M\pi_\rel}$, large $\theta_\e$
requires a lens that is either massive or nearby,
both of which suggest that it is bright.  On the other
hand, a small $\theta_*$ implies that the source is faint.
Generally, if a faint source and a bright potential lens
are close on the sky, only the lens will be seen, until
it starts to strongly magnify the source.  This has important
implications for the real time recognition of extreme
magnification events, as we discuss in \S~\ref{sec:summary}.
Here we review the evidence as to whether the blended light
in OGLE-2005-BLG-343 is in fact the lens.

{\hei
As was true for OGLE-2003-BLG-175/MOA-2003-BLG-45 mentioned in 
\S~\ref{sec:lense}, the blended light in OGLE-2004-BLG-343 
lies in the ``reddening sequence'' of foreground
disk stars.}  It is certainly ``dazzling'' by any criterion, being
about 50 times brighter than the source in $I$ and 150 times
brighter in $V$ (see Fig.~\ref{fig:fcmd}).
Is the blended light also due to the lens in this case?

There is one argument for this hypothesis and another against.
We initiate the first by estimating the mass and distance to the
blend as follows.
We model the extinction due to dust at a distance $x$ along the line of
sight by $d A_I/dx = 0.4\,\kpc^{-1} e^{-qx}$ and set $q=0.26\,\kpc^{-1}$
in order to reproduce the measured extinction to the bulge
$A_I(8\,\kpc) = 1.34$.  Using the \citet{reid91} color-magnitude
relation, we then adjust the distance to the blend until it
reproduces the observed color and magnitude of the blend.
We find a distance modulus of 12.6 ($\sim3.3\,\kpc$), and with the aid of the
\citet{cox} mass-luminosity relation, we estimate a corresponding
mass $M_{\rm l} = 0.9\,\msun$.  Inspecting Figure~\ref{fig:fmcs},
we see that this is almost exactly the peak of the lens-distance
distribution function predicted by combining light-curve
information and the Galactic model.  This is quite striking because,
in the absence of light-curve information, the lens would be
expected to be relatively close to the source. {\hei
From our Monte-Carlo simulation toward the line of sight
of this event, the total prior probability of 
the bulge-bulge events is about 1.5 times higher than the prior 
probability of the bulge-disk events, and furthermore, 
only about $7\%$ of all events have lenses less than $3.3\,\kpc$ away 
(see green and purple histograms in source and lens 
distance modulus panels of Fig.~\ref{fig:fmcs}).}
 It is only because the
light curve lacks obvious finite-source effects
(despite its very high-magnification) that one is forced to consider
lenses with large $\theta_\e$, which generally drives one toward
nearby lenses in the foreground disk. Based on our experience analyzing
many blended microlensing events, the blended light is most
often from a bulge star rather than a disk star, which simply
reflects the higher density of bulge stars. In brief, it is quite
unusual for lenses to be constrained to lie in the disk, and it
is quite unusual for events to be blended with foreground disk
stars.
This doubly unusual set of circumstances would be more
easily explained if the blend were the lens.

However, if the blend were the lens, then the source and lens
would be aligned to better than 1 mas during the event, and
one would therefore expect that
the apparent position of the source would not change as the
source first brightened and then faded.  In fact, we find that
the apparent position does change by about $73\pm 9\,$mas.
However, since the apparent source (i.e., combined source and blended
light at baseline) has a near neighbor at 830 mas,
which is almost as bright as the source/blend, it is quite possible
that the lens actually is the blend, but that this neighbor is
corrupting the astrometry.

Thus, the issue cannot be definitively settled at present.
However, it could be resolved in principle by, for example,
obtaining
high-resolution images of the field a decade after the event when
the source and lens have separated sufficiently to both be seen.
If the blend is the lens, then they will be seen moving directly
apart with a proper motion given $\mu=\theta_\e/t_\e$, where
$\theta_\e$ is derived from  the estimated mass and distance to the
lens
and $t_\e$ is the event timescale.

Since the blend cannot be positively identified as the lens, we
report our main results using a purely probabilistic estimate
of the lens parameters.  However, for completeness, we also report
results here based on the assumption that the lens is the blend.
{\hei Compared to the previous simulation, in which we considered the full
mass function and full range of distances, we sample only the 
narrow intervals of mass and distance that are consistent with the
observed color and magnitude of the lens/blend. To implement
these restrictions}, we repeat the Monte Carlo, but with the additional
constraint that the predicted apparent magnitudes agree with the
observed blend magnitude (with an error of 0.5 mag) and that the
predicted
colors (using the above extinction law and the \citealt{reid91}
color-magnitude relation) also show good agreement 
with the observed color (with 0.2 mag error).
These errors are, of course, much larger than the observational
errors.  They are included to reflect the fact that the theoretical
predictions for color and magnitude at a given mass are not
absolutely accurate.

\begin{figure}
\plotone{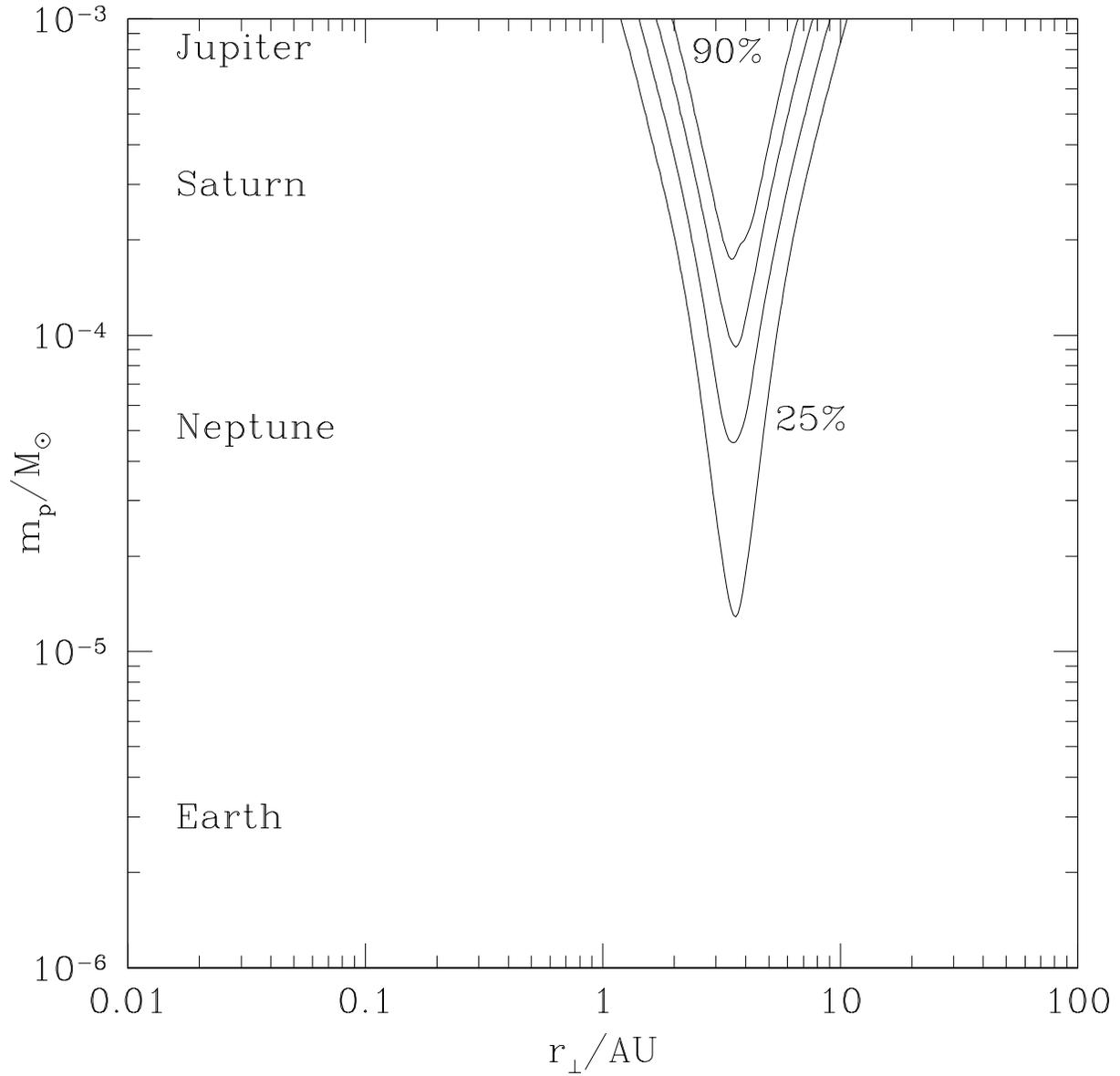}
\caption{
\label{fig:lensblend}
Planetary detection efficiency as a function of $r_{\perp}$, the physical 
projected star-planet distance, and $m_p$, the planetary mass for 
OGLE-2004-BLG-343, by assuming that the blended light is due to the lens. 
The contours represent 25\%, 50\%, 75\%, and 90\% efficiency.}
\end{figure}

Figure~\ref{fig:lensblend} is the resulting version of
Figure~\ref{fig:mrreal}
when the Monte Carlo is constrained to reproduce the blend color and
magnitude.  The sensitivity contours are narrower and deeper,
reflecting
the fact that the diagram no longer averages over a broad range of lens
masses but rather is restricted effectively to a single mass (and
single distance).

\section{Summary and Discussion
\label{sec:summary}}
In this paper we present our analysis of microlensing event OGLE-2004-BLG-343, with the 
highest peak magnification ($A_{\rm max}=3000 \pm 1100$) ever analyzed to date. The light curve
is consistent with the single-lens microlensing model, and no planet has been detected in this event.
We demonstrate that if the peak had been well covered by the observations, the event would have had 
the best sensitivity to planets to date, and it would even have had some sensitivity to Earth-mass 
planets (\S~\ref{sec:resfake}, \S~\ref{sec:realspe}). 
However, this potential has not been fully realized due to human error (\S~\ref{sec:data}), and
OGLE-2004-BLG-343 turns out to be no more sensitive to planets than a few other 
high-magnification events analyzed before (\S~\ref{sec:res}, \S~\ref{sec:realspe}). 
Thus, while ground-based microlensing surveys are technically sufficient 
to detect very low-mass planets, the relatively short timescale of the sensitive regime of 
high-magnification microlensing events demands a rapidity of response that
is not consistently being achieved. In the final paragraph below, we develop several suggestions to 
rectify this situation.

In \S~\ref{sec:model} we show that finite-source effects are important in analyzing this event,
so we extend the method of \citet{yoo423} to incorporate such effects in planetary detection 
efficiency analysis. Moreover, since magnification calculations of
binary-lens models with finite-source effects are computationally remarkably expensive, and applying previous 
finite-source algorithms, it would have taken of order a year of CPU time to do the detection efficiency 
calculation required by this event. We therefore develop two new
binary-lens finite-source algorithms (Appendix~\ref{sec:code}) that are considerably more efficient 
than previous ones. The ``map-making'' method (Appendix~\ref{sec:mapmake}) is an improvement on the 
conventional inverse ray-shooting method, which proves to be especially efficient for use
in detection efficiency calculations, while the ``loop-linking'' method (Appendix~\ref{sec:looplink}) 
is more versatile and could be easily implemented in programs aimed at finding best-fit 
finite-source binary-lens solutions.
{\hei Using these algorithms, we were able to complete the computations for
this paper in about 4 processor-weeks, roughly an order of magnitude
faster than would have been required using previous algorithms.}

Finally, we show in \S~\ref{sec:luminouslens} that the blend, which is a Galactic disk star, 
might very possibly be the lens, and that this case also proves to be highly 
probable from the Monte-Carlo simulation. However, it seems to contradict the astrometric evidence,
 and we point out that this issue could in principle be solved by future high-resolution 
images. Among the high-magnification events discovered by current microlensing survey 
groups, it is very likely that the lens star, which is also the apparent source, 
of those events is in the Galactic disk. Thus the blended light is usually far brighter 
than the source, thereby increasing the difficulty in early identification
 of such events. This fact motivates the first of several suggestions aimed at improving the 
recognition of very high-magnification events:\begin{itemize}
\item[1)] When events are initially alerted they should be accompanied by instrumental 
CMD of the surrounding field,
with the location of the apparent ``source'' highlighted.  Events
whose apparent sources lie on the ``reddening sequence'' of foreground
disk stars (see Fig.~\ref{fig:fcmd}) have a high probability to actually
be lenses of more distant (and fainter) bulge sources.  These
events deserve special attention even if their initial light curves
appear prosaic. \item[2)] For each such event it is possible to
measure the {\it color} (but not immediately the magnitude) of the
source by the standard technique of obtaining two-band photometry
and measuring the slope of the relative fluxes in the two bands.
If the color is different from that of the apparent ``source''
at baseline, that will prove that this baseline light is not
primarily due to the source, and it will increase the probability that this baseline
object is the lens.  Moreover, if the source color is relatively red, it will
show that the source is probably faint and so is (1) most likely
{\it already} fairly highly magnified (thereby making it
possible to detect above the foreground blended light) and (2) capable, 
potentially at least, of being magnified to very high
magnification (see \S~\ref{sec:luminouslens}).  This would motivate obtaining
more data while the event was still faint to help predict its
future behavior and would enable a guess as to how to ``renormalize''
the event's apparent magnification to its true magnification.
This is important because generally one cannot accurately determine
this renormalization until the event is within 0.4 mag (when the event
is $t_{\rm eff}$ before its peak), at which point it may well be too 
late to act on this knowledge. \item[3)] Both survey groups and follow-up 
groups should issue alerts on suspected high-magnification events guided 
by a relatively low threshold of confidence, recognizing that this will 
lead to more ``false alerts'' than at present.  If such alerts are 
accompanied by a cautionary note, they will promote intergroup discussions 
that could lead to more rapid identification of high-magnification events
without compromising the credibility of the group.
\end{itemize}

\acknowledgments

We would like to thank Jaiyul Yoo and Dale Fields for their generous help.
We thank Phil Yock, Ian Bond, Bohdan Paczy\'nski and Juna Kollmeier 
for their insightful comments on the manuscript.
S.\,D.\ and A.\,G.\ were supported by NSF grant AST 042758.
D.\,D., A.\,G., and R.\,P.\ were supported by NASA grant NNG04GL51G.
B.\,S.\,G.\ was supported by a Menzel Fellowship from the Harvard College
Observatory.
C.\,H.\ was supported by the SRC program of Korea Science \&
Engineering Foundation.
B.-G.\,P.\ acknowledges support from the Korea Astronomy and
Space Science Institute.
Support for OGLE was provided by Polish
MNII grant 2P03D02124, NSF grant AST-0204908, and NASA grant
NAG5-12212. A.\,U.\ acknowledges support from the grant ``Subsydium
Profesorskie'' of the Foundation for Polish Science. 
This publication makes use of data products from the Two Micron All Sky Survey, 
which is a joint project of the University of Massachusetts and the Infrared 
Processing and Analysis Center/California Institute of Technology, 
funded by the National Aeronautics and Space Administration and the National Science Foundation.
Any opinions, findings, and conclusions or recommendations
expressed in this material are those of the authors and
do not necessarily reflect the views of the NSF.

\appendix
\section{Two New Finite-Source Algorithms}
\label{sec:code}

To model planetary light curves,  we develop two new binary-lens
finite-source algorithms.  The first algorithm, called ``map-making'',
is the main work horse.  For a fixed $(b,q)$ geometry, it can
successfully evaluate the finite-source magnification of
almost all data points on the light curve and can robustly
identify those points for which it fails.  The second algorithm,
called ``loop-linking'', is much less efficient than map-making
but is entirely robust.  We use loop-linking whenever the map-making
routine decides it cannot robustly evaluate the magnification of
a point.  In addition, at the present time, map-making does
not work for resonant lensing geometries, i.e., geometries for
which the caustic has six cusps.  For planetary
mass ratios, resonant lensing occurs when the planet is very 
close to the Einstein radius, $b\sim 1$.  We use loop-linking
in these cases also.

\subsection{Map-Making
\label{sec:mapmake}}

Map-making has two components: a core function that evaluates the
magnification and a set of auxiliary functions that test whether
the measurement is being made accurately.  If a light-curve point fails
these tests, it is sent to loop-linking.

Finite-source effects are important when the source passes over or
close to a caustic.  Otherwise, the magnification can be evaluated
using the point-source approximation, which is many orders of magnitude
faster than finite-source evaluations.  Hence, the main control issue
is to ensure that any point that falls close to a caustic is evaluated
using a finite-source algorithm or at least is tested to determine
whether this is necessary.  For very high magnification events, the
peak points will always pass close to the central caustic.  Hence,
the core function of map-making is to ``map'' an Einstein-ring
annulus in the image plane that covers essentially all of the possible
images of sources that come close to the central caustic.  The method
must also take account of the planetary caustic(s), but we address that
problem further below.

We begin by inverse ray-shooting an annulus defined by
$A_{\rm PSPL} > A_{\rm min}$, where $A_{\rm PSPL}$ is the 
\citet{pac86} magnification due to a point source by a point lens and
$A_{\rm min}$ is a suitably chosen threshold.  For OGLE-2000-BLG-343,
we find that $A_{\rm min}=75$ covers the caustic-approaching points in
essentially all cases.  The choice of the density of the ray-shooting
map is described below.  Each such ``shot'' results in a four-element
vector $(x_i,y_i,x_s,y_s)$.  We divide the portion of the source
plane covered by this map into a rectangular grid with $k=1, \ldots ,N_g$
elements.  We choose the size of each element to be equal to the
smallest source radius being evaluated by the map.  Hence, each
``shot'' is assigned to some definite grid element $k(x_s,y_s)$.  We
then sort the ``shots'' by $k$.  For each light-curve point to be
evaluated, we first find the grid elements that overlap the source.
We then read sequentially through the sorted 
file\footnote{Whether this ``file'' should actually be an external
disk file or an array in internal memory depends on both the size
of the available internal memory and the total number of points
evaluated in each lens geometry.  In our case, we used internal
arrays.} from the beginning
of the element's ``shots'' to the end.  For each ``shot'', we ask
whether its $(x_s,y_s)$ lies within the source.  If so, we weight
that point by the limb-darkened profile of the source at that radius.
Note that a source with arbitrary shape and surface brightness profile
would be done just as easily.

For each light-curve point, we first determine whether at least one
of the images of the center of the source lies in the annulus.
In practice, for our case, the points satisfying this condition
are just those on the night of the peak, but for other events
this would have to be determined on a point-by-point basis.  
We divide those points with images outside the annulus into two classes,
depending on whether they lie inside or outside two or three rectangles
that we ``draw'', one around each caustic.  Each rectangle is larger
than the maximum extent of the caustics by a factor of $1.5$ in each direction.
If the source center lies outside all of these rectangles, we assume
that the point-source approximation applies and evaluate the magnification
accordingly.  If it lies inside one of the rectangles (and so either
near or inside one of the caustics), we perform the following test
to see whether the point-source approximation holds.  We evaluate the
point-source magnifications at five positions, namely, the source center
$A(0,0)$, two positions along the source $x$-axis $A(\pm \lambda\rho_*,0)$,
and two positions along the source y-axis $A(0,\pm \lambda\rho_*)$, where
$\lambda\leq 1$ is a parameter.
We demand that
\begin{equation}
\label{eqn:gauditest}
\bigg|{A(\lambda\rho_*,0)+A(-\lambda\rho_*,0)\over 2A(0,0)}-1\bigg|+
\bigg|{A(0,\lambda\rho_*)+A(0,-\lambda\rho_*)\over 2A(0,0)}-1\bigg|<4\sigma,
\end{equation}
where $\sigma$ is the maximum permitted error 
(defined in \S~\ref{sec:algparms}).
We use a minimum of five values in order to ensure that the magnification 
pattern interior to the source is reasonably well sampled; 
the precise value is chosen empirically and is a compromise between 
computing speed and accuracy.  We require the number of values
to be at least $2[\rho_*/\sqrt{q}]$ in order to ensure that small, 
well-localized perturbations interior to the source caused by low
 mass ratio companions are not missed.

If a point passes this test, the magnification pattern in the
neighborhood of the point is adequately represented by a gradient
and so the point-source approximation holds.  Points failing this
test are sent to loop linking.

The remaining points, those with at least one image center lying in the
annulus, are almost all evaluated using the sorted grid as described above.
However, we must ensure that the annulus really covers all of the images.
We conduct several tests to this end.

First, we demand that no more than one of the three or five images of the source
center lie outside the annulus.  In a binary lens, there is usually
one image that is associated with the companion and that is highly
demagnified.  Hence, it can generally be ignored, so the fact that
it falls outside the annulus does not present a problem.  If more
than one image center lies outside the annulus, the point is sent to
loop-linking.
Second, it is possible that an image of the center of the source could
lie inside the annulus, but the corresponding image of another point
on the source lies outside.  In this case, there would be some
intermediate point that lay directly on the boundary.  To guard against
this possibility, we mark the ``shots'' lying within one grid step
of the boundaries of the annulus,
and if any of these boundary ``shots'' fall in the source, we send
the point to loop-linking.
Finally, it is possible that even though the center of the source lies
outside the caustic (and so has only three images), there are other parts
of the source that lie inside the caustic and so have two additional
images.  If these images lay {\it entirely} outside the annulus, the
previous checks would fail.  However, of necessity, some of these
source points lie directly on the caustic, and so their images lie
directly on the critical curve.  Hence, as long as the critical
curve is entirely covered by the annulus, at least some of each of
these two new images will lie inside the annulus and the ``boundary
test'' just mentioned can robustly determine whether any of these
images extend outside the annulus.
For each $(b,q)$ geometry, we directly check whether the annulus covers 
the critical curve associated
with the central caustic by evaluating the critical curve locus
 using the algorithm of \citet{witt90}.

\subsection{Loop-Linking
\label{sec:looplink}}

Loop-linking is a hybrid of two methods: inverse ray-shooting and 
Stokes's theorem.  In the first method (which was also used above in
``map-making''), one finds the source location corresponding 
to each point in the image plane.  Those that fall inside the
source are counted (and weighted according to the local surface
brightness), while those that land outside the source are not.
The main shortcoming of inverse ray-shooting is that one must
ensure that the ensemble of ``shots'' actually covers the entire image 
of the source without covering so much additional ``blank space''
that the method becomes computationally unwieldy.

In the Stokes's theorem approach, one maps the boundary of
(a polygon-approximation of) the source into the image plane, 
which for a binary lens yields either three or five closed polygons.
These image polygons form the (interior or exterior) boundaries
of one to five images.  If one assumes uniform surface brightness,
the ratio of the combined areas of these images (which can
be evaluated using Stokes's theorem) to the area of the source
polygon is the magnification.  

There are two principal problems with the Stokes's theorem approach.
First, sources generally cannot be approximated as having
uniform surface brightness.  This problem can be resolved simply
by breaking the source into a set of annuli, each of which is
reasonably approximated as having uniform surface brightness.
However, this multiplies the computation time by the number of
annuli.  Second, there can be numerical problems of several types
if the source boundary passes over or close to a cusp.  First,
the lens solver, which returns the image positions given the
source position, can simply fail in these regions.  This at
least has the advantage that one can recognize that there
is a problem and perhaps try some neighboring points.  The
second problem is that even though the boundary of the source
passes directly over a cusp, it is possible that none of the vertices 
of its polygon approximation lie within the caustic.  The polygonal
image boundary will then fail to surround the two new images of
the source that arise inside the caustic, so these will not be
included in the area of the image.  Various steps can be taken to
mitigate this problem, but the problem is most severe for very
low-mass planets (which are of the greatest interest in the present
context), so complete elimination of this problem is really an uphill
battle.

The basic idea of loop-linking is to map a polygon that is {\it slightly
larger} than the source onto the image plane, and then to inverse
ray shoot the interior regions of the resulting image-plane polygons.
This minimizes the image-plane region to be shot compared to other
inverse-ray shooting techniques.  It is, of course, more time-consuming
than the standard Stokes's theorem technique, but it can accommodate
arbitrary surface-brightness profiles and is
more robust.  As we detail below, loop-linking can fail at any
of several steps.  However, these failures are always recognizable,
and recovery from them is always possible simply by repeating the
procedure with a slightly larger source polygon.  

Following \citet{gg97}, 
the vertices of the source polygon are each mapped to an array of
three or five image positions, each with an associated parity.  If the lens solver
fails to return three or five image positions, the evaluation is repeated
beginning with a larger source polygon.  For each successive pair
of arrays, we ``link'' the closest pair of images that has the
same parity and repeat this process until all images in these
two arrays are linked.  An exception occurs when one array
has three images and the other has five images, in which case two images
are left unmatched.  Repeating this procedure
for all successive pairs of arrays produces a set of linked ``strands'',
each with either positive or negative parity.  The first element
of positive-parity strands and the last element of negative-element
strands are labeled ``beginnings'' and the others are labeled ``ends''.
Then the closest ``beginning'' and ``end'' images are linked and
this process is repeated until all ``beginnings'' and ``ends''
are exhausted.  The result is a set of two to five linked loops.
As with the standard Stokes's theorem approach, it is possible
that a source-polygon edge crosses a cusp without either
vertex being inside the caustic.  Then the corresponding image-polygon
edge would pass inside the image of the source, which would cause
us to underestimate the magnification.  We check for this possibility
by inverse ray shooting the image-polygon boundary (sampled with
the same linear density as we later sample the images) back
into the source plane.  If any of these points land in the source,
we restart the calculation with a larger source polygon.

We then use these looped links to efficiently locate the region
in the image plane to do inverse ray shooting.  We first
examine all of the links to find the largest difference, $\Delta y_{\rm max}$,
between the y-coordinates of the two vertices of any link.  Next, we
sort the $m=1,\ldots ,n$ 
links by the lower y-coordinate of their two vertices, $y_m^-$.
One then knows that the upper vertex obeys 
$y_m^+ \leq y_m^- + \Delta y_{\rm max}$.  Hence, for each 
$y$-value of the inverse ray shooting grid, we know that only
links with $y_m^-\leq y \leq y_m^- + \Delta{y_{\rm max}}$ can
intersect this value.  These links can quickly be identified
by reading through the sorted list from $y_m^-=y-\Delta y_{\rm max}$ 
to $y_m^-=y$.  The $x$ value of each of these crossing links
is easily evaluated.  Successive pairs of $x$'s then bracket the
regions (at this value of $y$) where inverse rays must be shot.
As a check, we demand that the first of each of these bracketing
links is an upward-going link and the second is a downward-going
link.

\subsection{Algorithm Parameters
\label{sec:algparms}}

Before implementing the two algorithms described above, one must first
specify values for certain parameters.  Both algorithms involve inverse
ray shooting and hence require that a sampling density by specified.
Let $g$ be the grid size in units of the Einstein radius.  For magnification
$A\gg 1$, the image can be crudely approximated as two long strands whose
total length is $\ell=4 A \rho_*$ and hence of mean width 
$(\pi\rho_*^2 A)/\ell = (\pi/4)\rho_*$.  If, for simplicity, 
we assume that the strand is aligned with the grid, then there will be 
a total of $\ell/g$ grid tracks running across the strand.  Each will
have two edges, and on each edge there will be an ``error'' of
$12^{-1/2}$ in the ``proper'' number of grid points due to the fact
that this number must be an integer, whereas the actual distance
across the strand is a real number.  Hence, the total number of grid
points will be in error by $[(2\ell/g)/12]^{1/2}$, while the total
number itself is $\pi(\rho_*/g)^2 A$.  This implies a fractional error
$\sigma$,
\begin{equation}
\label{eqn:griderr}
\sigma^{-2} = {[\pi(\rho_*/g)^2 A]^2\over (2\ell/g)/12}
= {3\pi^2\over 2} A(\rho_*/g)^3.
\end{equation}
In fact, the error will be slightly smaller than given by equation
(\ref{eqn:griderr}) in part because the strand is not aligned
with the grid, so the total number of tracks across the strands
will be lower than $2\ell/g$, and in part because the ``discretization
errors'' at the boundary take place on limb-darkened parts of the
star, which have lower surface brightness, so fluctuations here have
lower impact.  
Hence, an upper limit to the grid size required to achieve a 
fractional error $\sigma$ is
\begin{equation}
\label{eqn:gridsize}
{g\over \rho_*} = \biggl({3\pi^2\over 2}\biggr)^{1/3}\sigma^{2/3} A^{1/3}.
\end{equation}

For each loop-linking point, we know the approximate magnification $A$
because we know the weighted parameters of the single-lens model.
We generally set $\sigma$ at 1/3 of the measurement error, so that the
(squared) numerical noise is an order of magnitude smaller than
that due to observational error.  Using equation (\ref{eqn:gridsize})
we can then determine the grid size.

For the map-making method, the situation is slightly more complicated.
Instead of evaluating one particular point as in the loop-linking
method, all of the points on the source plane with the same $(b,q,u_0,\rho)$
are evaluated within one map. Therefore, the grid size for this map
is the minimum value from equation~(\ref{eqn:gridsize}) to achieve the 
required accuracy for all of these points. We derive the following equation from
equation~(\ref{eqn:gridsize}) to determine the grid size in the map-making method:
\begin{equation}
\label{eqn:gridsize-map}
g = \biggl[{3\pi^2\over2}{{Q(u_0)}\over{u_0}}\biggl]^{1/3}\rho_*
\end{equation}
where
\begin{equation}
\label{eqn:qu0}
Q(u_0)=min_{A_i>75}
\Biggl\{{{{F(t_i)-F_{\rm b}(u_0)}\over{F_{\max}(t_i)-F_b(u_0)}}\sigma_i^2}\Biggl\}
\end{equation}

We find that $Q(u_0)=2.65\times10^{-7}$ is independent of $u_0$
for both the observational and simulated data of OGLE 2004-BLG-343. In principle, one could
determine a minimum $g$ for all $(u_0,\rho)$ combinations and generate only one map 
for a given $(b,q)$ geometry, but this would render the calculation unnecessarily long 
for most $(u_0,\rho)$ combinations. Instead we evaluate $g$ for each $(u_0,\rho)$ pair
and create several maps, one for each ensemble of $(u_0,\rho)$ pairs with similar $g$'s.
The sizes of the {\hei ensembles} should be set to minimize the total time spent generating, loading
and employing maps. Hence, they will vary depending on the application.

\section{MOA-2003-BLG-32/OGLE-2003-BLG-219}
\label{sec:moa}

\begin{figure}
\plotone{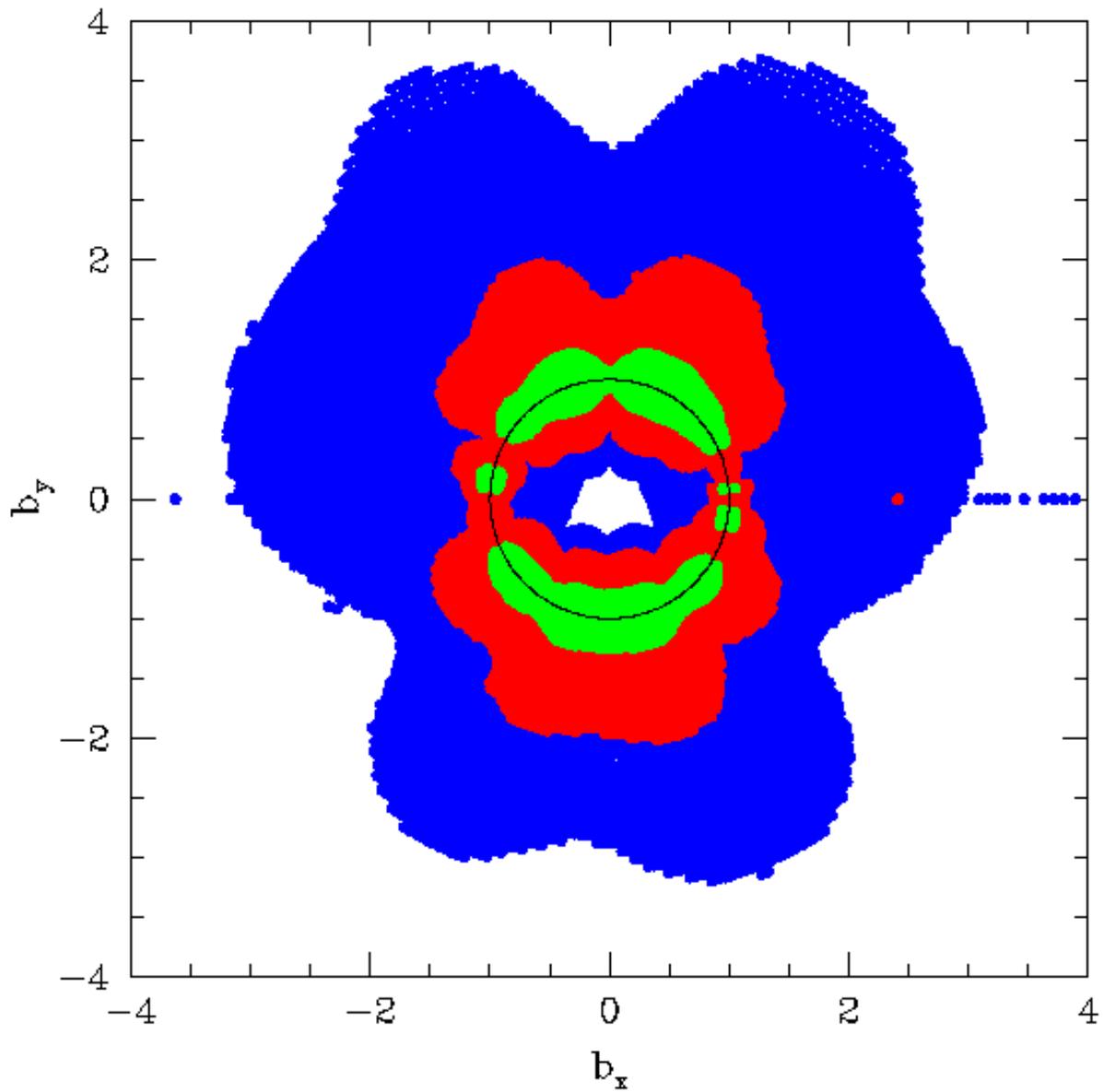}
\caption{
\label{fig:m32}
Planetary exclusion regions for microlensing event MOA-2003-32/OGLE-2003-BLG-219
as a function of projected coordinate $b_x$ and $b_y$ 
at planet-star mass ratio 
$q=10^{-5}$ ({\it green}), $10^{-4}$ ({\it red}) and $10^{-3}$ ({\it blue}). 
The source size (normalized to ${\theta}_\e$) $\rho_*$ 
is equal to 0.0007. The black circle is the Einstein ring, i.e., $b=1$.}
\end{figure}

MOA-2003-BLG-32/OGLE-2003-BLG-219, with a peak magnification $A_{\rm max}=525\pm 75$
is most sensitive to low-mass planets to date \citep{abe04}. However, 
instead of fitting the simulated binary-lens light curves to single-lens
models, \citet{abe04} obtain their $\Delta{{\chi}^2}$ by directly subtracting 
the ${\chi^2}$ of a simulated binary-lens light curve from that of the light curve 
that is the best fit to the data. 
Since the source star of this event could reside in the Sagittarius 
dwarf galaxy, which makes the Galactic modeling rather complicated, 
we do not attempt to apply our entire method to this event. 
We calculate planet exclusion regions 
with the same $\Delta{{\chi}^2}$ thresholds (60 for $q=10^{-3}$ and 
40 for the $q<10^{-3}$) 
as \citet{abe04} but using our method of obtaining $\Delta{{\chi}^2}$ 
by fitting the simulated binary-lens light curves to single-lens models. 
Figure~\ref{fig:m32} shows our results for the exclusion regions at 
planet-star mass ratios $q=10^{-5}, 10^{-4},$ and $10^{-3}$ for 
MOA-2003-BLG-32/OGLE-2003-BLG-219. The exclusion region we have obtained at 
$q= 10^{-3}$ is about $1/4$ in vertical direction and $1/9$ in horizontal 
direction relative to the corresponding region in \citet{abe04}, 
and the size of our exclusion region at $10^{-4}$ is 
about 60\% in each dimension relative to that in \citet{abe04}. 
Although according to our analysis, \citet{abe04} overestimate the sensitivity of 
MOA-2003-BLG-32/OGLE-2003-BLG-219 to both Jupiter-mass and Neptune-mass planets, 
their estimates of sensitivity to Earth-mass planets are basically consistent 
with our results and MOA-2003-BLG-32/OGLE-2003-BLG-219 nevertheless retains the 
best sensitivity to planets to date.

\newpage
\begin{deluxetable}{cccccccc}
\tablecaption{OGLE-2004-BLG-343 Best-Fit PSPL Model Parameters}
\tablewidth{0pt}
\tabletypesize{\scriptsize}
\tablehead{
 \colhead{$t_0$(HJD')} 
& \colhead{$u_0$}
& \colhead{$t_\e$(days)} 
& \colhead{$I_s$} 
& \colhead{$I_b$} 
& \colhead{$\chi^2$}
& \colhead{Degree of Freedom}
}
\startdata
$3175.7467 \pm 0.0005$ & 
$0.000333 \pm 0.000121$&
$42.5 \pm 15.6$        &
$22.24 \pm 0.40$    &
$18.08 \pm 0.01$    &
200.1 &
188 &
\\
\enddata
\label{table1}
\end{deluxetable}

\end{document}